\documentclass[sigconf]{acmart}
\usepackage{pifont}
\usepackage{graphicx}  
\usepackage{float}  
\usepackage{subfigure} 
\usepackage{algorithm}
\usepackage{algorithmicx}
\usepackage{algpseudocode}
\usepackage{amsmath}

\usepackage{enumerate}
\usepackage{enumitem}
\usepackage{multirow}
\usepackage{balance}

\AtBeginDocument{%
  \providecommand\BibTeX{{%
    \normalfont B\kern-0.5em{\scshape i\kern-0.25em b}\kern-0.8em\TeX}}}

\copyrightyear{2023}
\acmYear{2023}
\setcopyright{rightsretained}
\acmConference[KDD '23]{Proceedings of the 29th ACM SIGKDD Conference on Knowledge Discovery and Data Mining}{August 6--10, 2023}{Long Beach, CA, USA}
\acmBooktitle{Proceedings of the 29th ACM SIGKDD Conference on Knowledge Discovery and Data Mining (KDD '23), August 6--10, 2023, Long Beach, CA, USA}
\acmDOI{10.1145/3580305.3599866}
\acmISBN{979-8-4007-0103-0/23/08}

\makeatletter
\gdef\@copyrightpermission{
 \begin{minipage}{0.3\columnwidth}
  \href{https://creativecommons.org/licenses/by/4.0/}{\includegraphics[width=0.90\textwidth]{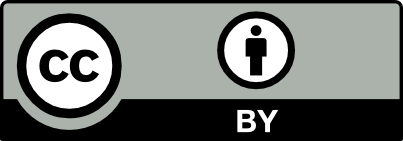}}
 \end{minipage}\hfill
 \begin{minipage}{0.7\columnwidth}
  \href{https://creativecommons.org/licenses/by/4.0/}{This work is licensed under a Creative Commons Attribution International 4.0 License.}
 \end{minipage}
 \vspace{5pt}
}
\makeatother

\begin{document}

\title{Modeling Dual Period-Varying Preferences for Takeaway Recommendation}



\author{Yuting Zhang}
\affiliation{%
  \institution{Institute of Computing Technology, \\
  Chinese Academy of Sciences}
  \city{Beijing}
  \country{China}}
\affiliation{
  \institution{University of Chinese Academy of Sciences}
  \city{Beijing}
  \country{China}}
\email{zhangyuting21s@ict.ac.cn}

\author{Yiqing Wu}

\affiliation{%
  \institution{Institute of Computing Technology, \\
  Chinese Academy of Sciences}
  \city{Beijing}
  \country{China}}
\affiliation{
  \institution{University of Chinese Academy of Sciences}
  \city{Beijing}
  \country{China}}
\email{wuyiqing20s@ict.ac.cn}

\author{Ran Le}
\affiliation{
  \institution{Meituan}
  \city{Beijing}
  \country{China}}
\email{leran@meituan.com}

\author{Yongchun Zhu}
\affiliation{
  \institution{Institute of Computing Technology, 
  Chinese Academy of Sciences}
  \city{Beijing}
  \country{China}}
\affiliation{
  \institution{University of Chinese Academy of Sciences}
  \city{Beijing}
  \country{China}}
\email{zhuyongchun18s@ict.ac.cn}

\author{Fuzhen Zhuang}
\authornote{Corresponding authors.}
\affiliation{%
  \institution{Institute of Artificial Intelligence, Beihang University}
  \city{Beijing}
  \country{China}}
  \affiliation{
      \institution{
        SKLSDE, School of Computer Science, Beihang University
      }
      \city{Beijing}
      \country{China}
  }
\email{zhuangfuzhen@buaa.edu.cn}

\author{Ruidong Han}
\affiliation{
  \institution{Meituan}
  \city{Beijing}
  \country{China}}
\email{hanruidong@meituan.com}

\author{Xiang Li}
\author{Wei Lin}
\affiliation{
  \institution{Unaffiliated}
  \city{Beijing}
  \country{China}}
\email{leo.lx007@qq.com}
\email{lwsaviola@163.com}

\author{Zhulin An}
\authornotemark[1]
\author{Yongjun Xu}
\affiliation{%
  \institution{Institute of Computing Technology, Chinese Academy of Sciences}
  \city{Beijing}
  \country{China}
}
\email{anzhulin@ict.ac.cn}
\email{xyj@ict.ac.cn}

\renewcommand{\shortauthors}{Yuting Zhang et al.}

\begin{abstract}
Takeaway recommender systems, which aim to accurately provide stores that offer foods meeting users' interests, have served billions of users in our daily life. Different from traditional recommendation, takeaway recommendation faces two main challenges: (1) \textbf{Dual Interaction-Aware Preference Modeling}. Traditional recommendation commonly focuses on users' single preferences for items while takeaway recommendation needs to comprehensively consider users' dual preferences for stores and foods. (2) \textbf{Period-Varying Preference Modeling}. Conventional recommendation generally models continuous changes in users' preferences from a session-level or day-level perspective. However, in practical takeaway systems, users' preferences vary significantly during the morning, noon, night, and late night periods of the day. To address these challenges, we propose a \textbf{D}ual \textbf{P}eriod-\textbf{V}arying \textbf{P}reference modeling (DPVP) for takeaway recommendation.  Specifically, we design a dual interaction-aware module, aiming to capture users' dual preferences based on their interactions with stores and foods. Moreover, to model various preferences in different time periods of the day, we propose a time-based decomposition module as well as a time-aware gating mechanism.  Extensive offline and online experiments demonstrate that our model outperforms state-of-the-art methods on real-world datasets and it is capable of modeling the dual period-varying preferences. Moreover, our model has been deployed online on Meituan Takeaway platform, leading to an average improvement in GMV (Gross Merchandise Value) of 0.70\%. 
\end{abstract}



\maketitle
\section{Introduction}
With convenient online ordering and offline delivery services,  takeaway platforms (e.g., Meituan\footnote{https://about.meituan.com/en}/Uber Eats\footnote{https://www.ubereats.com}) are playing an increasingly important role in our daily life. The total number of users in takeaway platforms has reached approximately 2.5 billion in 2022~\cite{now_statistic}. 
Aiming to accurately provide stores with foods that meet users' interests from a large store candidate pool,  takeaway recommender systems generally capture users’  preferences for foods or stores through their historical interaction behaviors with stores or foods. Compared with traditional recommendation, takeaway recommendation appears greatly more difficult due to the complicated relationships among users, stores and foods. Thus, takeaway recommender systems have attracted great attention from the research community and industry field ~\cite{yang2022gated,qi2021trilateral,yu2021dual,wang2020calendar,lin2022spatiotemporal}.

\begin{figure}
    \centering
    \includegraphics[width=\linewidth]{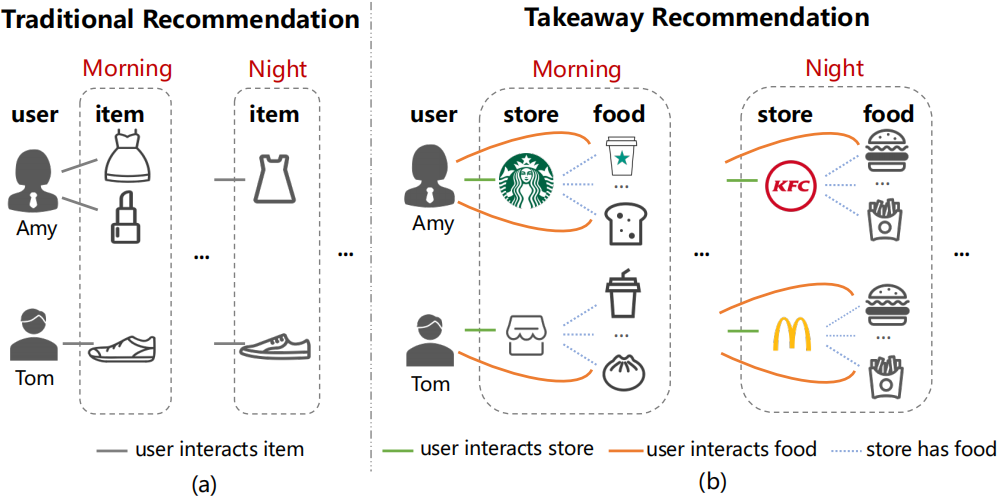}
    \caption{Illustration of differences between traditional recommendation and takeaway recommendation.}
    \label{fig:interation}
\end{figure}

Generally, traditional recommendation~\cite{he2017neural,zheng2017joint,seo2017representation,chen2018neural,zhang2019deep} models users' period-agnostic preferences from the single historical interaction data between users and items as Figure \ref{fig:interation}(a) shown.  However, the following two notable characteristics of takeaway systems make it challenging to directly apply traditional methods for takeaway recommendation.

(1) \textbf{Dual Interaction-Aware Preference}. As shown in Figure~\ref{fig:interation}, traditional recommendation, such as e-commerce recommendation, usually model user-item simple binary interaction for users' preferences for items, while the interaction between users, stores, and foods in takeaway recommendation appears more complex including three types of interaction relations. In particular, users' dual interactions with stores and foods imply users' preferences at different levels. Additionally, with the auxiliary relations between stores and foods, there is a more complex interplay between the users' dual preferences. For example, Tom might be fond of food baozi instead of caring about the store, while Amy might pay more attention to the brand of the dine-in store. Also, some people may weigh both preferences for the store and its food information, such as making a balance between high store ratings and low food prices. 

\begin{figure}
    \centering
    \includegraphics[width=0.58\linewidth]{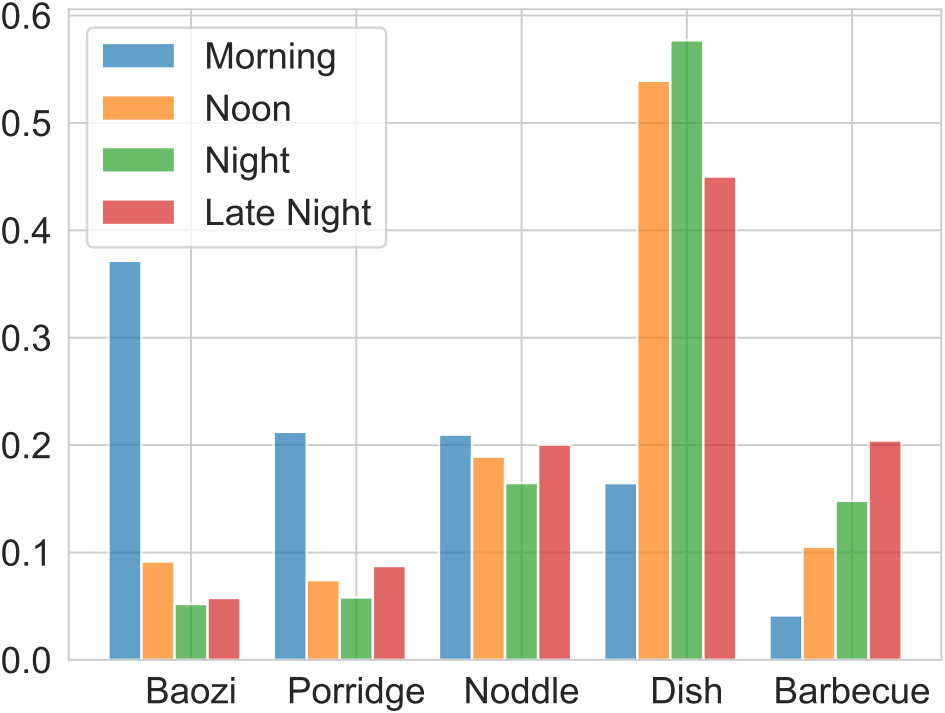}
    \caption{Proportion distribution of clicks between users and five major food categories in four different time periods. Note that the sum of the proportions of the five food categories in each time period is equal to 1.}
    \label{fig:distribution_time_interation}
\end{figure}

(2) \textbf{Period-Varying Preference}. 
As depicted in Figure~\ref{fig:interation}, in traditional recommendation, there is no obvious feature of user preferences changing with the time period of the day (i.e., period-agnostic preference modeling), while in takeaway recommendation, user preferences, especially for food, change significantly in different periods of a day. 
Visually, we obtain a total of 100,000 click interactions between users and five major food categories in a real takeaway dataset, and further present the proportion distribution of clicks on different categories of food in each period (i.e., Morning/Noon/Night/Late Night) according to the click occurring time in Figure \ref{fig:distribution_time_interation}. Note that as a traditional Chinese food, baozi generally serves as breakfast in the Morning.  Apparently, in the Morning period,  users are greatly more likely to interact with baozi than barbecue, while the opposite is true during Late Night. Particularly, user interactions in different periods also imply similar preferences. For example, when a user prefers spicy food, he/she is inclined to interact with spicy-flavored food over different periods.

Existing takeaway recommendation methods commonly leverage the coarse-grained single interaction data between the user and store or food and may further utilize the time as auxiliary information \cite{yang2022gated,qi2021trilateral,lin2022spatiotemporal,yu2021dual}. With feature-enriched user and store representations as input, ~\cite{yu2021dual} utilized the classical two-tower model to make store recommendations based on the historical interaction data between users and stores. Further, ~\cite{lin2022spatiotemporal} adopted an attention-based fusion mechanism at different times for users' single-level time-enhanced preference for foods, which is incapable of modeling users’ period-varying preferences within a day.  In brief, the existing methods have not systematically modeled users' dual period-varying preferences for food and store.

To address the above issues, we propose a novel model, named DPVP, for takeaway recommendation.  Specifically, we design a dual interaction-aware module, aiming to capture users' preferences based on the ternary interaction data shown in Figure ~\ref{fig:interation}(b). Furthermore, We propose a time-based decomposition module to capture users' dual-period-specific preferences in the period hierarchy and simultaneously learn users' period-shared preferences. Then, a time-aware gate network is utilized to fuse the user's dual period-specific and period-shared preferences to obtain the user's dual period-varying preferences. Eventually,  the model make a prediction based on the balance between the users' dual period-varying preferences for stores and foods. 
To sum up, our contributions are listed below:
\begin{itemize}[parsep=0pt] 
    \item We highlight the challenges of dual interaction-aware and time-varying preference modeling in takeaway recommendation, which is more complicated than traditional recommendation modeling, and further design a novel model to address these challenges. To the best of our knowledge, we are the first to systematically model dual time-varying preferences for takeaway systems.
    \item We propose a dual interaction-aware module for capturing users' dual preferences. Besides, we design a time-based decomposition to model users' period-specific preferences in the period hierarchy, and further design a time-aware gate to adaptively model users' period-varying preferences.
    \item Both offline and online experiments were conducted to show the effectiveness of our model and the great ability of the model to capture the dual period-varying preferences for takeaway recommendation. 
\end{itemize}

\section{Related work}
\subsection{Takeaway Recommendation}
In the takeaway recommender system, multiple interactions of user, store, and food lead to complex relationships among the three, which is different from the simple binary interaction relationship usually modeled in traditional recommender systems~\cite{he2017neural,zheng2017joint,seo2017representation,chen2018neural,zhang2019deep}. 
For example, on the basis of the assumption that users with similar behaviors have similar preferences on stores, traditional Collaborative Filtering (CF)~\cite{ekstrand2011collaborative} has been extensively applied in modern takeaway recommender systems. Since CF-based methods~\cite{he2017neural,koren2009matrix,rendle2009bpr} suffer from data sparsity and cold start issues, many works tried to utilize additional information such as time information~\cite{yang2022gated,wang2020calendar,qi2021trilateral,lin2022spatiotemporal}, spatial information\cite{yang2022gated,qi2021trilateral,lin2022spatiotemporal} and attributes~\cite{yu2021dual,wang2020calendar,lin2022spatiotemporal} to enhance recommendation accuracy. For example, DAT~\cite{yu2021dual} enriched the user and store representations based on the corresponding user features and store attribute information for the recommendation.  
Moreover, TRISAN~\cite{qi2021trilateral} adopted an attention-based fusion mechanism to capture the rich interplay between spatial information and time information to improve recommendation performance. Further, considering the importance of  time information in takeaway systems, some methods attempted to not simply utilize the time information as the auxiliary input of the designed model. For example, AutoIntent ~\cite{li2022automatically} treated the interaction time factors as vertex nodes in graphs and then extracted the high-order relations of three groups of hypergraphs among user preferences for the food category factor, time factor, and spatial factor respectively for the recommendation. 
However, existing takeaway recommender systems rarely involve the inherent ternary complex user-store-food relationships and period-varying characteristics in the systems.

\subsection{Graph-based Recommendation}
With the rapid development and huge success of Graph Neural Network (GNN) in various machine learning areas~\cite{zhou2020graph,zhang2020deep,wu2020comprehensive,hamilton2017inductive}, great efforts have been devoted to applying it to the recommendation~\cite{wu2022graph,eksombatchai2018pixie,pfadler2020billion,ying2018graph,li2022automatically}. Most data in recommender systems essentially implies graph structure ~\cite{berg2017graph, ying2018graph}.  For example, the interaction data can naturally form a bipartite graph between user and item nodes, with observed interactions denoted by links between corresponding nodes. Given the graph data, GNN mainly focuses on iteratively aggregating information from neighbors and integrating the aggregated information with the current central node representation during the propagation process.
For example, GC-MC~\cite{berg2017graph} aggregated the information from neighbors based on the types of ratings.  Standard GCN~\cite{kipf2016semi} factorized user-item rating matrices into user and item embedding matrices for recommendation.
NGCF~\cite{wang2019neural} advanced GCN by additionally encoding the interactions via an element-wise multiplication. 
Due to the discovery of  the non-necessity of nonlinear activation and feature transformation in GCN~\cite{wu2019simplifying}, SGCN~\cite{wu2019simplifying} and LightGCN\cite{he2020lightgcn} were proposed for recommendation by removing these two parts. Moreover, owning to the powerful representation ability of GNN, a great number of GNN-based methods ~\cite{ying2018graph,zhao2019intentgc,wu2019dual,xu2020graphsail}  have been deployed in the industry to produce high-quality recommendation results. For example, Pinsage~\cite{ying2018graph} combined efficient random walks and graph convolutions to generate embeddings of items that incorporate both graph structure as well as item feature information, which was the first work applying GCN in industrial recommender systems.
DANSER~\cite{wu2019dual} took advantage of the interaction between users and articles and social relations for a real-world article recommender system, WeChat Top Story. 
Different from previous works, our method retains the traditional user–item structure,  extending it to dual period-varying tripartite graphs among users, stores and foods, aiming at modeling the users' dual interaction-aware and period-varying preferences for takeaway recommendation.    


\section{Preliminaries}

In practical takeaway systems, an interaction commonly forms ``who interacts (e.g., clicks) which food in which store at what time", which involves a user, a store, several foods as well as a time period. To comprehensively model the interactions, we define the \textit{period-varying ternary interaction data} in this work as follows:

\textbf{Definition 1}.\textbf{\textit{Period-Varying Ternary Interaction Data}}. Given ${\mathcal{U},\mathcal{S},\mathcal{O},\mathcal{M}}$ denoting the universal sets of users, stores, foods, and time periods, each record $x$ in the interaction dataset $\mathcal{X}$ can be formulated as $x=(u,s,O,m)$, which represents that user $u \in \mathcal{U}$ interacts $N$ foods $O=\{o_1,o_2,\cdots,o_N\}$ (where $o_j \in \mathcal{O},\forall j=1,2,...,N$)  in store $s \in \mathcal{S}$ at time period $m \in \mathcal{M}$.

Specifically, we divide a day into 4 time periods: Morning, Noon, Night and Late Night, i.e., $M=4$ with $M$ denoting the number of time periods $|\mathcal{M}|$. Example interaction records can be found in Figure~\ref{fig:interation}(b). Based on the ternary interaction data, the \textit{full-period global graph} including user-store, user-food and store-food interactions can be defined as follows: 

\textbf{Definition 2}.\textbf{\textit{Full-Period Global Graph}}. 
Given the ternary interaction dataset $\mathcal{X}$, the \textit{full-period global graph} can be formulated as $\mathcal{G}=\{\mathcal{U}\cup\mathcal{S}\cup\mathcal{O},\mathcal{{E}_{US}}\cup \mathcal{{E}_{UO}} \cup \mathcal{{E}_{SO}\}}$.  Let $\mathcal{{E}_{US}},\mathcal{{E}_{UO}}$  and $\mathcal{{E}_{SO}}$ be edges that represent user–store, user–food, store–food relations. For each record $x=(u,s,{O},m)$, there exists the nodes $\{u,s,o_j\}$ and the bi-directional interaction edges $(u,s) \in \mathcal{{E}_{US}},(u,o_j) \in \mathcal{{E}_{UO}},(s,o_j) \in \mathcal{{E}_{SO}}$. Each edge is attached with a time period attribute $m$. 

\begin{figure*}
    \centering
    \includegraphics[width=\linewidth]{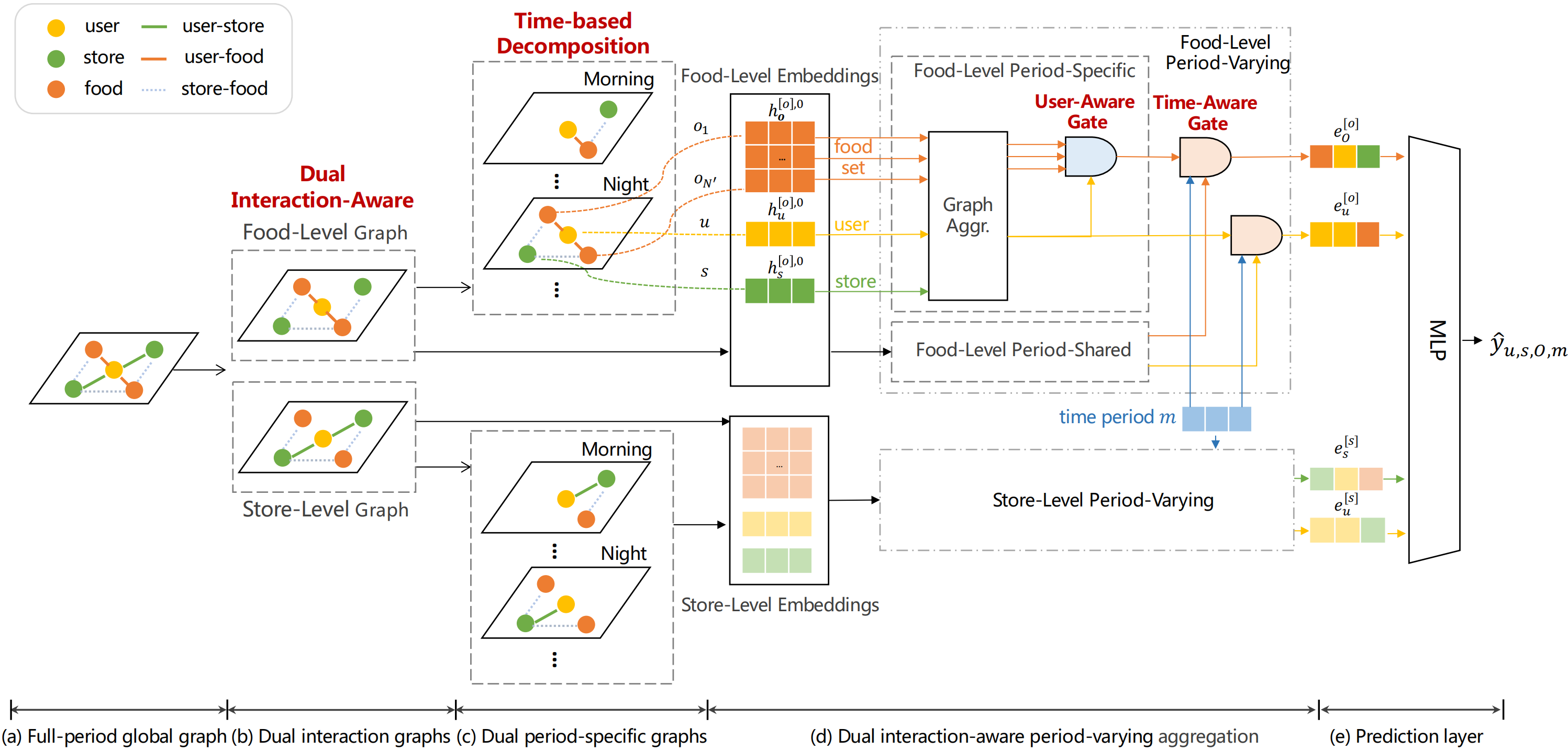}
    \caption{Overview of our proposed DPVP model.}
    \label{fig:framework}
    
\end{figure*}

Particularly, $\mathcal{G}$ is depicted in Figure ~\ref{fig:framework}(a).
In this work, we only focus on recommending stores, an urgent practical task for takeaway applications, while the food recommendation is left as future work. In order to consider the user's preference for food in the store recommendation, we define the most frequently clicked food set of the store $s$ with the fixed size $N'$ as $\Gamma (s)$. Based on the above definitions, the takeaway recommendation task can be defined as follows:

\textbf{Definition 3}.\textbf{\textit{ Takeaway Recommendation}}. Given the \textit{period-varying ternary interaction data} $\mathcal{X}$ and \textit{full-period global graph} $\mathcal{G}$, DPVP aims to recommend the stores $s \in \mathcal{S}$ with the food set $\Gamma (s)$ that user $u$ would be interested in at time period $m$.

\section{Methodology} 
In this section, we will illustrate in detail how to realize the aforementioned takeaway recommendation based on the \textit{period-varying ternary interaction data} and \textit{full-period global graph} constructed based on it. The overview of our proposed DPVP is depicted in Figure~\ref{fig:framework}.
Firstly, we propose an interaction-aware module to decompose the \textit{full-period global graph} into store-level and food-level graphs, which can also be denoted as dual interaction graphs. Then, we design a unified time-based decomposition module to deal with the dual graphs to obtain dual period-specific graphs.  After aggregation on dual full-period and period-specific graphs, our model can model users' dual period-shared and period-specific preference representations. In particular, due to the multiple foods in the food set, a user-aware gate mechanism is designed to effectively ensemble different food information to obtain the personalized representation of the food set.  Then, a time-aware gate network is designed to adaptively obtain the dual period-varying representations, finally followed by a prediction layer for our takeaway recommendation.

In the following, we first describe the construction process of dual period-varying graphs including the dual interaction-aware and time-based decomposition modules in Section~\ref{sec:graph_construction}.  Then, we detailedly illustrate the aggregation of dual interaction-aware graphs for enriching the representations as well as the user-aware gating network for personalized food representation in Section~\ref{sec:dual interaction-aware}. Furthermore, we present the modeling of the period-varying preferences in Section ~\ref{sec:period-varying}, followed by the prediction layer and optimization process illustrated in Section~\ref{sec:prediction}. 


\subsection{Dual Period-Varying Multigraphs Construction}\label{sec:graph_construction}
\subsubsection{Dual Interaction-Aware Module}
To capture the dual preferences, we propose an interaction-aware module to decompose \textit{full-period global graph} into the dual interaction graphs, i.e., food-level and store-level graphs shown in Figure~\ref{fig:framework}(b). 
Specifically, in the food-level graph with food as the central node,  there exist two kinds of interaction edges: (1) user-food interaction edge implies the user's preference for food; (2) food-store edge which can form a food-store-food structure indicating the same-store relationship between foods. Similarly, in the store-level graph with the store as the central node,  there are user-store and store-food edges,  where the formed store-food-store structure implies the food similarity of stores.  Thereby, the dual interaction graphs including food-level and store-level graphs can be correspondingly formulated as:
\begin{equation}
    \begin{aligned}
    \mathcal{G}^{[o]} &=\{\mathcal{U}\cup\mathcal{S}\cup\mathcal{O},\mathcal{{E}_{UO}} \cup \mathcal{{E}_{SO}}\},\\
    \mathcal{G}^{[s]}&=\{\mathcal{U}\cup\mathcal{S}\cup\mathcal{O},\mathcal{{E}_{US}}\cup \mathcal{{E}_{SO}}\}.   
\end{aligned}
\end{equation}

\subsubsection{Time-Based Decomposition Module}
With the aim of learning the period-varying preferences, we design a unified time-based decomposition module to deal with the dual interaction graphs. Specifically, each edge set in dual graphs is unified as $\mathcal{E}$, and the time period  of each interaction edge $c$ is denoted as $c_t$. In the time-based decomposition module, the hard weight of each edge $c$ and the edge set in time period $m$ can be formulated as:
\begin{equation}
    \begin{aligned}
      w_c^m &= \mathbb{I}(c_t=m),\\
      \mathcal{{E}}^m &= \{w_c^m*c \}, \forall c \in \mathcal{E},
    \end{aligned}
\end{equation}
where $\mathbb{I}(\cdot)$ represents the indicator function. Thereby, given the time period $m$ of each interaction data, the dual interaction graphs $\mathcal{G}^{[o]},\mathcal{G}^{[s]}$ can be decomposed into $M$ dual period-specific graphs depicted in Figure ~\ref{fig:framework}(c) as $\forall m=1,2,... ,M$:
\begin{equation}
    \begin{aligned}
\mathcal{G}_{m}^{[o]}&=\{\mathcal{U}\cup\mathcal{S}\cup\mathcal{O},\mathcal{{E}}^m_\mathcal{UO}\cup \mathcal{{E}}^m_\mathcal{SO}\}, \\
\mathcal{G}_{m}^{[s]}&=\{\mathcal{U}\cup\mathcal{S}\cup\mathcal{O},\mathcal{{E}}^m_\mathcal{US}\cup \mathcal{{E}}^m_\mathcal{SO}\}.
    \end{aligned}
\end{equation} 

\subsection{Dual Interaction-Aware Preference}\label{sec:dual interaction-aware}
\subsubsection{Dual Embedding Layers}\label{sec:dual_embedding}
Before aggregating graphs, we propose to initialize the embeddings of users, stores, and foods. Since we hope to capture the user's preferences for two completely different types of entities (i.e., store and food) in dual subgraphs,  we initialize two groups of embeddings for the two graphs.  Moreover, intuitively, users hold the aforementioned period-shared preferences in different time periods and there is no significant difference in the meaning of foods and stores over different time periods. Therefore, we utilize the same initialization embedding group of the same interaction-aware level for the subgraphs of different time periods at the same food/store level.

Take a user $u \in \mathcal{U}$ as an example: after turning $u$ into the food-level user matrix $H_{u}^{[o]} \in \mathbb{R}^{|\mathcal{U}|\times d }$ and the store-level matrix $H_{u}^{[s]} \in \mathbb{R}^{|\mathcal{U}|\times d}$ with $d$ denoting the embedding dimension, we can obtain the initialization embedding of user node $u$ as $h_{u}^{[o],0}$ and $h_{u}^{[s],0}$ respectively.
For the subgraphs of time period $m$, the user embeddings of the food-level and store-level are initialized as $h_{u,m}^{[o],0} =h_{u}^{[o],0}$ and  $h_{u,m}^{[s],0} = h_{u}^{[s],0}$ correspondingly. 
Since the dual interaction-aware preference modeling process of dual graphs in all periods and period-specific graphs is the same and the aggregation method of the food-level graphs is similar to that of store-level  graphs,  we mainly elaborate the food-level period-specific preference modeling process of graphs $\mathcal{G}_{m}^{[o]}$ in time period $m$. 

\subsubsection{Single-Level Preference}
Following~\cite{hamilton2017inductive,sun2019multi},  we simply apply the mean-pooling operation to aggregate neighbor embeddings in the food-level graph $\mathcal{G}_{m}^{[o]}$. 
In detail, the aggregated neighbor embedding for the central food node in the food-level graph can be formulated as:
\begin{equation}
    \begin{aligned}
      \mathbf{h}_{o,{m}}^{[o],l+1} = \frac{1}{c_{{o,m}}^{(u,o)}} \sum\limits_{(u,o) \in \mathcal{{E}}^m_\mathcal{UO}}\mathbf{h}_{u,{m}}^{{[o]},l}
      +\frac{1}{c_{o,m}^{(s,o)}} \sum\limits_{(s,o) \in \mathcal{{E}}^m_\mathcal{SO}}\mathbf{h}_{s,{m}}^{{[o]},l},
    \end{aligned}
\end{equation}
where $\frac{1}{c_{{o,m}}^{(u,o)}}$ and $\frac{1}{c_{{o,m}}^{(s,o)}}$ are normalization constants and ${c_{{m}}^{(u,o)}}/{c_{{m}}^{(s,o)}}$ is equal to the in-degree from neighbor user/store nodes to the food node of $o$. 
Since both the user and store interact solely with food nodes in the food-level graph, the aggregated neighbor representations of the user and store can be formulated as:
\begin{equation}
    \begin{aligned}
   \mathbf{h}_{u,{m}}^{[o],l+1} &= \frac{1}{c_{{u,m}}^{(u,o)}} \sum\limits_{(u,o) \in \mathcal{{E}}^m_\mathcal{UO}}\mathbf{h}_{o,{m}}^{{[o]},l},\\
      \mathbf{h}_{s,{m}}^{[o],l+1} &= \frac{1}{c_{{s,m}}^{(s,o)}} \sum\limits_{(s,o) \in \mathcal{{E}}^m_\mathcal{SO}}\mathbf{h}_{o,{m}}^{{[o]},l}.
    \end{aligned}
\end{equation}

Eventually, the weighted-pooling operation is applied to generate the aggregated user/food  representations in the food-level graph at time period $m$  by operating on propagated $L$ layers:
\begin{equation}
    \begin{aligned}
    \mathbf{h}_{o,{m}}^{[o],*} = \sum\limits_{l=0}^{L} \alpha_l \mathbf{h}_{o,{m}}^{[o],l},  \quad
    \mathbf{h}_{u,{m}}^{[o],*} = \sum\limits_{l=0}^{L} \alpha_l \mathbf{h}_{u,{m}}^{[o],l}, 
    \end{aligned}
\end{equation}
where $\alpha_l$  denotes the importance of the $l$-th layer representation in constituting the final embedding. Similar to LightGCN~\cite{he2020lightgcn}, we uniformly set $\alpha_l$ as $1/(l+1)$ since the focus of our work is not on the choice of $\alpha_l$.
Therefore, after the aggregation of the food-level graph, we can obtain the food-level representations of the user $u$ and food $o$ as $\mathbf{h}_{o,{m}}^{[o],*}$ and $\mathbf{h}_{u,{m}}^{[o],*}$.
After the similar aggregation method, we can also obtain the final aggregated embeddings of user $u$ and store $s$ in store-level graph as $\mathbf{h}_{u,{m}}^{[s],*},\mathbf{h}_{s,{m}}^{[s],*}$.

\subsubsection{Personalized Food Representation}
For each store, its $N'$ most frequently clicked foods are selected as its candidate foods. However, users' interest in food varies from person to person. For instance, user A may prefer baozi, while user B is more interested in noodles in the same store. In order to model the personalized food representation, a user-aware gate mechanism is designed to effectively ensemble different food information in the candidate food set $O=\Gamma (s)$.
Specifically, the user-aware gating network $g_{u}$ produces a distribution over $N'$ foods  based on the input, and the final ensembled personalized food representation is formulated as the weighted sum of the graph representations of all the candidate foods of the store:

\begin{equation}\label{eq:food gate}
    \begin{aligned}
    g_{u}(o_i) &= \frac{exp(\mathbf{h}_{o_i,m}^{[o],*} \times {{\mathbf{h}_{u,{m}}^{[o],*}}}^\top)}{\sum\limits_{j=1}^{N'} \mathop{exp}({\mathbf{h}_{o_j,m}^{[o],*} \times {{\mathbf{h}_{u,{m}}^{[o],*}}}^\top})},\\
    \mathbf{h}_{{{O}},m}^{[o],*} &= \sum\limits_{i=1}^{N'} g_{u}(o_i)\mathbf{h}_{o_i,m}^{[o],*}.
    \end{aligned}
\end{equation}


\subsection{Period-Varying Preference}\label{sec:period-varying}
To capture the period-varying patterns of the interaction, we obtain the period-specific enriched graph embedding group at each time period $m$ as $\mathbf{h}_{u,m}^{[o],*},\mathbf{h}_{{O},m}^{[o],*},\mathbf{h}_{u,m}^{[s],*},\mathbf{h}_{s,m}^{[s],*}$. Ideally, when making a recommendation in time period $m$,  DPVP ought to merely utilize the corresponding graph embedding group in time period $m$ for prediction. In other words, this is a hard one-hot integration method. However, considering the following factors: (1) the user's interactions in different periods imply similar preferences; (2) period-specific data appear sparse compared to full-period interaction data, we adopt a soft time-aware gating method instead of the hard integration method for the integration of representations in different time periods. The soft time-aware gate combines the representations in different time periods and full time periods, and further filters out irrelevant information while strengthening the relevant information of the target time period.
 Without loss of generality, due to the same integration methods of representations $\mathbf{h}_{u,m}^{[o],*}$,$\mathbf{h}_{O,m}^{[o],*}$,$\mathbf{h}_{u,m}^{[s],*}$,$\mathbf{h}_{s,m}^{[s],*}$, we take $\mathbf{h}_{u,m}^{[o],*}$ as an example to illustrate the time-aware gate integration as follows:
\begin{equation}
    \begin{aligned}
    g_{m}(u,m) &= \frac{exp(\mathbf{h}_{u,m}^{[o],*} \times {[\mathbf{h}_{u}^{[o],*}, \mathbf{e}_m]}^\top)}{\sum\limits_{k=1}^M exp(\mathbf{h}_{u,k}^{[o],*} \times {[\mathbf{h}_{u}^{[o],*}, \mathbf{e}_k]^\top)}},\\
    \mathbf{e}_{u,m}^{[o]} &= \sum\limits_{m=1}^M g_{m}(u,m)\mathbf{h}_{u,m}^{[o],*},
    \end{aligned}
\end{equation}
where $[\cdot]$ denotes the operation of concatenation and $\mathbf{e}_{m}$ indicates the embedding of time period $m$.

\subsection{Prediction and  Optimization}\label{sec:prediction}
After the above operations, we obtain the dual period-varying representations. That is, $\mathbf{e}_{u,m}^{[o]},\mathbf{e}_{o,m}^{[o]}$ generated from food-level graphs and $ \mathbf{e}_{u,m}^{[s]},\mathbf{e}_{s,m}^{[s]}$ aggregated from store-level graphs. Since the candidate food set ${O}$ contains the fine-grained food information of the store $s$, we employ MLP (Multi-Layer Perceptrons) to build a bridge between fine-grained food and coarse-grained store levels to predict how likely the user would adopt the store. 
Eventually, the predicted score of candidate store $s$ with candidate food set ${O}$ for user $u$ at time period $m$ can be formulated as:  
\begin{equation}
    \begin{aligned}\label{eq:predict}
    \hat{y}_{u,s,{O},m}&=MLP([\mathbf{e}_{u,m}^{[o]},\mathbf{e}_{O,m}^{[o]},\mathbf{e}_{u,m}^{[s]},\mathbf{e}_{s,m}^{[s]}]).
    \end{aligned}
\end{equation}


Then, we adopt the Bayesian Personalized Ranking (BPR)~\cite{rendle2009bpr} loss to emphasize that the observed interaction should be assigned a higher score than unobserved ones as follows:
\begin{equation}
    \begin{aligned}
    \mathcal{L} = \sum\limits_{(u,s,s')\in {Y},{O} =\Gamma (s),{O}' =\Gamma (s')}-\ln \sigma(\hat{y}_{u,s,{O},m}-\hat{y}_{u,s',{O}',m}),
    \end{aligned}\end{equation}
where ${Y}={(u,s,s')|(u,s)\in {R}^+,(u,s')\in {R}^-}$ denotes the pairwise training data; ${R}^+$ indicates the positive observed interaction set, and ${R}^-$ represents the randomly-sampled negative set; $\sigma(\cdot)$ stands for the sigmoid function.

\section{Experiments}
In this section, we present empirical results to demonstrate the effectiveness of our proposed DPVP. These experiments are designed to answer the following research questions:
\begin{itemize}
    \item \textbf{RQ1} How does DPVP perform compared with state-of-the-art recommendation models?
    \item \textbf{RQ2} What is the effect of dual interaction-aware preference and period-varying preference modeling in our proposed DPVP?
    \item \textbf{RQ3} How does DPVP perform in the real-world online recommendations with practical metrics?
    \item \textbf{RQ4} How do hyper-parameters in DPVP impact recommendation performance?

\end{itemize}
\subsection{Experimental Settings}
\textbf{Dataset Description}.
We conduct experiments on two real-world datasets\footnote{We collected the Meituan datasets because no public datasets are suitable for our task. We will release part of  experimental datasets later and our code will be available in https://github.com/17231087/DPVP.git.} (denoted as MT-large and MT-small datasets, with large differences in data size) of two cities in two different time intervals from the Meituan Takeaway platform, one of the largest takeaway platforms in China. The time span of both datasets is eight days.
In order to model users' period-varying preferences, we group the whole day into four time periods, which are 5: 00-10: 00 for Morning, 10: 00-15: 00 for Noon, 15:00-20:00 for Night and 20:00-5:00+1 for Late Night. To evaluate the model performance, we split the first six days’ data for training, the following one-day data for validation, and the last day’s data for testing. 
For each ground truth test data, we randomly sample 99 stores that user did not interact with as negative samples. Detailed statistics of these two datasets and detailed implementation are presented in Appendix~\ref{appendix:data_description} and ~\ref{appendix:implementation}.  

\textbf{Evaluation Metrics.}
To evaluate the performance of the recommendation models, we adopt four commonly-used metrics including three ranking metrics MRR\cite{radev2002evaluating}, NDCG@K~\cite{jarvelin2002cumulated} and Hit@K (we set K as 10 by default), and an accuracy metric AUC\cite{ferri2011coherent}. 

\begin{table*}[]
    \centering
    \begin{tabular}{c|cccc|cccc}
    \toprule
    \multirow{2}{*}{Model} & \multicolumn{4}{c|}{MT-small} & \multicolumn{4}{c}{MT-large}\\
    \cline{2-9}
    & Hit@10 &  NDCG@10 &AUC & MRR  &Hit@10 & NDCG@10 &AUC & MRR \\
    \midrule
    
    NeuMF &0.3957 &0.2104 &0.7814 &0.1788&0.4554&0.2488&0.8198&0.2101 \\
    
    DNN&0.4190&0.2503&0.7549&0.2188&0.6073 &0.3898 & 0.7903&0.3174 \\ 
    ENMF&0.5513&0.3845&0.7356&\underline{0.3450}&0.5911&\underline{0.4285}&0.7496&\underline{0.3887}\\
    SimpleX&0.4851&0.2710&0.8171&0.2276&0.5634&0.3289&0.8457&0.2763\\
    \midrule
    GCN &0.4446&0.2478&0.7968&0.2110&0.6288&0.4199&0.8535&0.3698 \\
    GAT &0.4384 &0.2488& 0.7529&0.2105 & 0.6202&0.3871&0.8361 &0.3290 \\
    NGCF &0.5009 &0.2839 & 0.8163&0.2411 & 0.6358&0.4222&0.8593 &0.3725\\
    HGT &0.4467 &0.2474 & 0.8032&0.2097 & 0.6217 &0.3905&0.8352 &0.3329   \\
    LightGCN  &0.4942  &0.2784&0.7979 &0.2330&0.6309&0.4250& 0.8562&0.3756\\
    Ultra-GCN &0.3666&0.2057&0.7624 &0.1777&0.4959& 0.3164&0.7751&0.2840\\
    
    SVD-GCN & \underline{0.5745}  & \underline{0.3973}&\underline{0.8205}& 0.3402 & \underline{0.6370} &0.4248&0.8609  &0.3736 \\
    DPVP(full-period global graph)  &0.5094 & 0.3022  &0.7875&0.2562 &0.6324   &0.4117&\underline{0.8621} &0.3592   \\
    \hline
    DPVP &\textbf{ 0.6167}*&\textbf{0.4180}*&\textbf{0.8392}*&\textbf{0.3715}* &\textbf{0.6599}* & \textbf{0.4563}*&\textbf{0.8741}*  & \textbf{ 0.4075}*\\
    Imp\% &+7.3455&+5.2102&+2.2791&+7.6812&+3.5950&+6.4877&+1.3919&+4.8366\\
    \bottomrule
    \end{tabular}
    \caption{Overall performance on MT-small and MT-large datasets. The last row Imp\%  indicates the relative improvements of the best performing method (bolded) over the strongest baselines (underlined) and marker * indicates that the improvement is statistically significant compared with the best baseline (paired t-test with p-value < 0.005).} 
    \label{tab:comparision_res}
\end{table*}
\begin{table*}[]
    \centering
    \begin{tabular}{c|cccc|cccc}
    \toprule
    
    \multirow{2}{*}{Model} & \multicolumn{4}{c|}{MT-small} & \multicolumn{4}{c}{MT-large}\\
    \cline{2-9}
     & Hit@10 &  NDCG@10 &AUC & MRR &Hit@10 &  NDCG@10 &AUC & MRR \\
    \midrule
    DPVP(user-food)&0.3990 &0.2531 &0.7487&0.2298& 0.4054 & 0.2299 &0.7651& 0.1986 \\
    DPVP(food-level) &0.4438 & 0.2750&0.7804& 0.2444&0.4547& 0.2842 &0.7845 & 0.2602  \\
    DPVP(user-store)& 0.5355  & 0.3432 &0.7888&0.3017 &0.6110 &0.3879&0.8558&0.3359\\
    DPVP(store-level) &0.5461&0.3588&0.7902&0.3161&0.6271 &0.4158 &0.8639  & 0.3664\\
    DPVP(global level) &0.5865  &0.3909 &0.8296 &0.3467 &0.6494  &0.4476 &0.8714 &0.3998  \\
   
    DPVP &\textbf{ 0.6167*}&\textbf{0.4180*}&\textbf{0.8392*}&\textbf{0.3715*} &\textbf{0.6599*} & \textbf{0.4563* }&\textbf{0.8741*}  & \textbf{ 0.4075*}\\
    \bottomrule
    \end{tabular}
    \caption{Impact of dual interaction-aware preference modeling and * indicates p-value < 0.005.}
    \label{tab:Dual Interaction-Aware}
\end{table*}
\begin{figure*}
    \centering
    \includegraphics[width=0.95\linewidth]{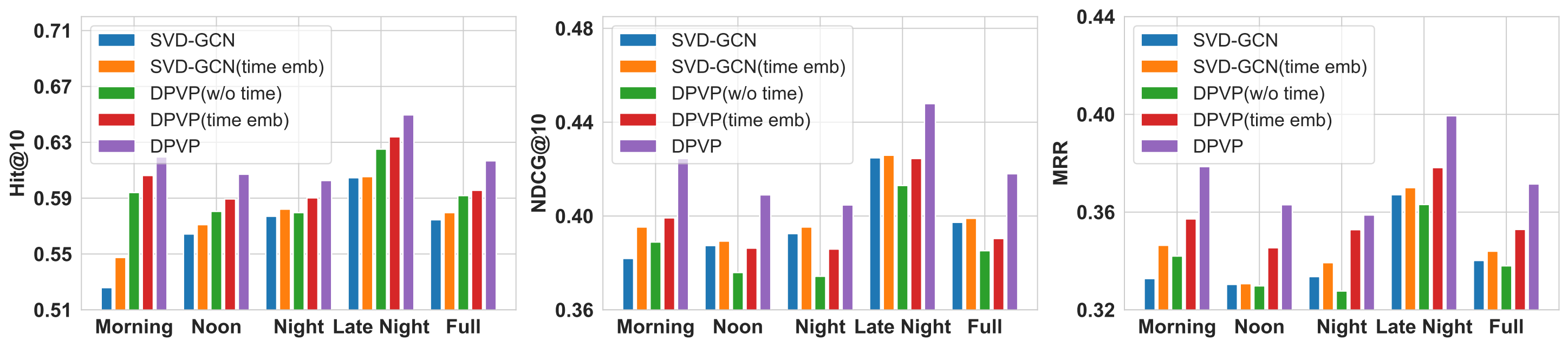}
    \caption{Impact of period-varying modeling on results for different time periods in MT-small dataset.} 
    \label{fig:small_time}
\end{figure*}
\begin{figure*}
    \centering
    \includegraphics[width=0.95\linewidth]{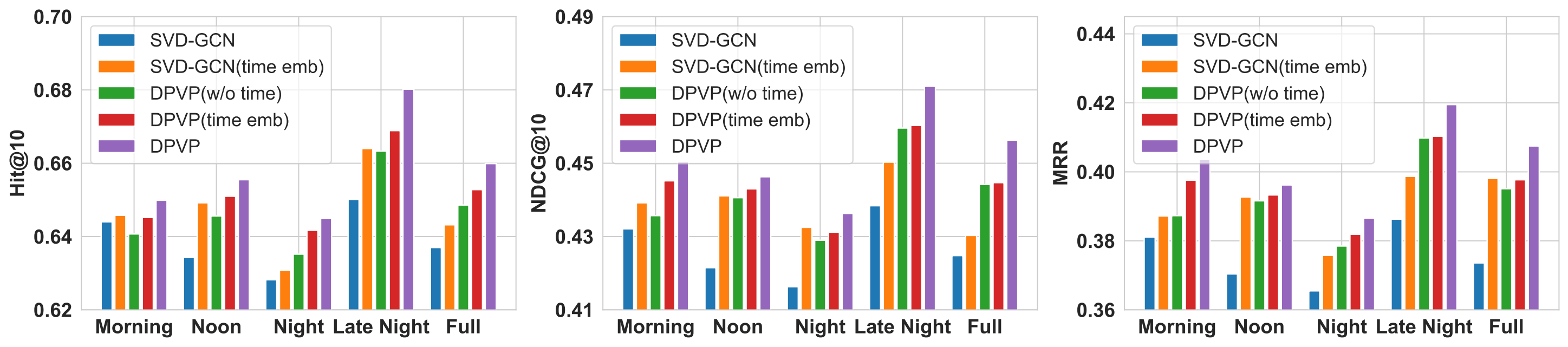}
    \caption{Impact of period-varying modeling on results for different time periods in MT-large dataset.}
    \label{fig:big_time}
\end{figure*}




\subsection{Overall Performance Comparison (RQ1)}\label{sec:overall_performance}
We compare our proposed models with four graph-free baselines (i.e., NeuMF~\cite{he2017neural}, DNN, ENMF~\cite{chen2020efficient}, SimpleX~\cite{mao2021simplex}) and seven graph-based baselines (i.e., GCN~\cite{scarselli2008graph}, GAT~\cite{velivckovic2017graph}, NGCF~\cite{wang2019neural}, HGT~\cite{hu2020heterogeneous}, LightGCN~\cite{he2020lightgcn}, UltraGCN~\cite{mao2021ultragcn} and SVD-GCN~\cite{peng2022svd}). The detailed information of these baselines is provided in Appendix~\ref{appendix:baselines}.
In particular, in order to test whether the improvement of the model performance is caused by the dual interaction-aware and period-varying preference modeling proposed in our DPVP rather than the graph aggregation method of DPVP, we design DPVP(full-period global graph) which makes prediction based on the \textit{full-period global graph} depicted in Figure ~\ref{fig:framework}(a). 
The overall performance on two adopted datasets is reported in Table ~\ref{tab:comparision_res}.
From the results, we can observe that:
\begin{itemize}
    \item DPVP significantly outperforms all the competitive baselines.  Compared to the best performance of baselines, for the two datasets, DPVP gains an average $5.86\%$ improvement in terms of the three ranking metrics as well as an average $1.84\%$ improvement with respect to AUC. Such significant gains verify that DPVP can greatly enhance the recommendation performance.
    \item The performance of DPVP(full-period global graph) with the same mean-pooling aggregation on the \textit{full-period global graph} is remarkably lower than that of DPVP. This observation demonstrates that the significant improvement of our model mainly lies in our proposed dual interaction-aware and period-varying preference modeling. 
    \item The graph-based methods universally outperform most graph-free methods, indicating the importance of enhancing interactions by incorporating high-order neighborhood information to alleviate data sparsity.
    \item The performance of MT-large with large data volume is better than that of MT-small with small data volume for the same model. The main reason would be that with more data in each time period, the model performs better in modeling time-related behaviors.

\end{itemize}

\begin{figure}
\centering
    \includegraphics[width=0.62\linewidth]{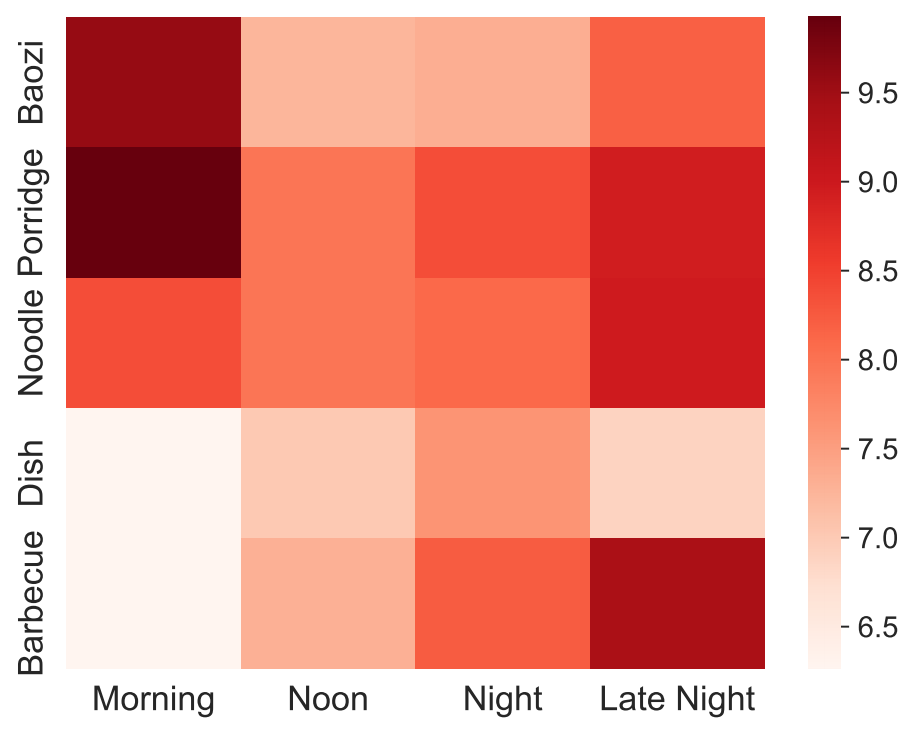}
    \caption{Visualization of the average predicted scores of five major food categories over different time periods.} 
    \label{fig:heat}
\end{figure}
\begin{figure*}
    \centering
    \includegraphics[width=0.95\linewidth]{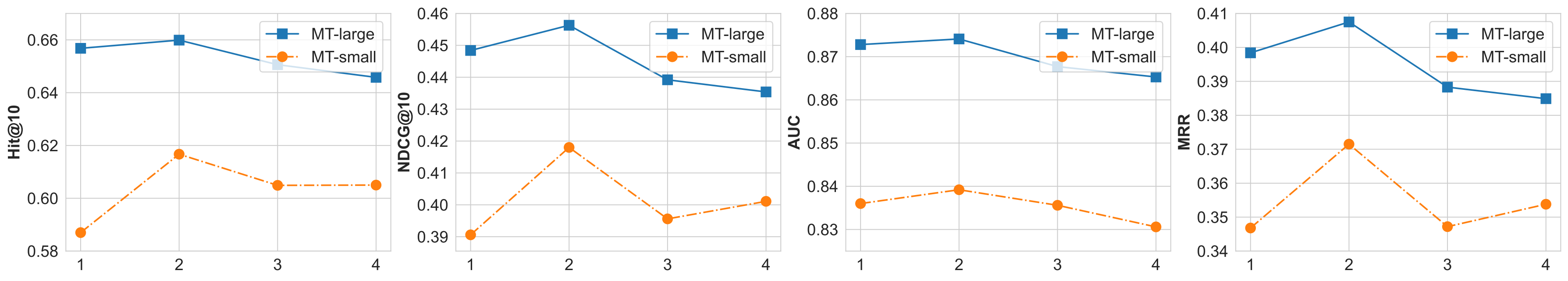}
    \caption{Performance  w.r.t
the number of layers $L$ on the two datasets.}
    \label{fig:layer_ab}
\end{figure*}
\begin{figure*}
    \centering
    \includegraphics[width=0.95\linewidth]{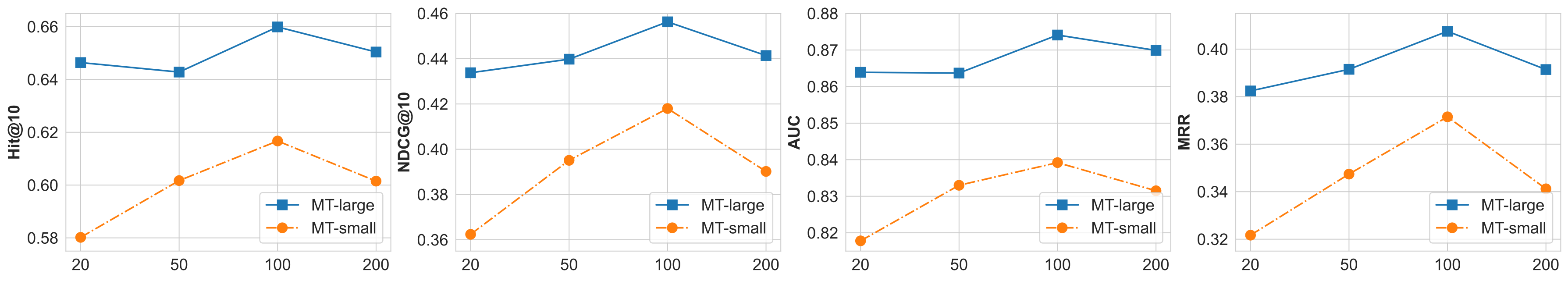}
    \caption{Performance  w.r.t
the dimension  of embedding  vectors $d$ on the two datasets.}
    \label{fig:dim_ab}
\end{figure*}

\subsection{Ablation Study (RQ2)}\label{sec:core_component}
As the modeling of dual interaction-aware and period-varying preferences are the core of DPVP, we conduct the following ablation studies to investigate the effectiveness of these core modules.
\subsubsection{Impact of Dual Interaction-Aware Preference} 
To investigate whether DPVP can benefit from dual interaction-aware preference modeling, we propose five ablation versions of DPVP: (1) DPVP (user-food). This baseline only considers the user-food interactions; (2) DPVP(user-store). This baseline merely considers the binary user-store interactions, usually modeled by the traditional recommender systems.
(3) DPVP(food-level). This baseline only retains food-level graphs by removing the store-level graphs in Figure ~\ref{fig:framework}(b); (4) DPVP(store-level). This baseline only retains store-level graphs by removing the food-level graphs depicted in Figure ~\ref{fig:framework}(b); (5) DPVP (global level). This baseline directly performs time-based decomposition on \textit{full-period global graph} without dual interaction-aware module, that is $\mathcal{G}_*^{[o]},\mathcal{G}_*^{[s]}$ are replaced with $\mathcal{G}_*$). The third and fourth baselines are devised to verify whether our method is still effective with only a single-level interaction graph.
For fairness, in order to avoid the influence of the parameter amount on the results, we double the embedding size of the single-level interaction and single-level graph model so that the parameter amount is the same as that of DPVP model.  Besides, for the global level model, we replace the dual aggregated user representations $\mathbf{e}_u^{[o]},\mathbf{e}_{u}^{[s]}$ with the global-level graph aggregated user representation respectively as part of the input of the prediction layer, so that the parameter amount is consistent with that of DPVP.
From the results shown in Table ~\ref{tab:Dual Interaction-Aware}, we have the following observations:
\begin{itemize}
    \item As we expected, compared with the models that only consider binary interaction information DPVP(user-food) and DPVP(user-store), the models that consider ternary interactive information has better performance.
    \item The performance of the single-level model is significantly worse than that of the dual interaction-aware model. This is within expectation because
    dual interaction-aware modeling can capture the additional level of user preferences compared to single-level modeling.
    \item Encouragingly, the performance of preference modeling based on dual interaction-aware graphs is better than that based on global-level graphs regardless of hierarchical modeling. This observation indicates the effectiveness of our proposed dual interaction-aware preference modeling. This result is not surprising because the global level graph models the interaction at the two levels at the same time, which may confuse the user's preferences at the two levels, ultimately resulting in poor user representation. On the contrary, our model explicitly captures the hierarchical preference at different levels by considering the interaction with users at different levels.  
    
\end{itemize}

\subsubsection{Impact of Period-Varying Modeling} 

In order to verify the effectiveness of period-varying preference modeling, we conduct three additional ablation baselines as follows:
\begin{itemize}
\item SVD-GCN(time emb). On the basis of the original SVD-GCN with only user and store input,  the time embedding $\mathbf{e}_m$ is added as an input to the prediction layer of SVD-GCN. 

\item DPVP(w/o time). This baseline completely removes the modeling of the period-specific subgraphs with only the dual interaction graphs in Figure~\ref{fig:framework}(b), which is designed to prove the effectiveness of period-varying modeling in DPVP.
\item DPVP(time emb). This baseline removes the modeling of the period-specific graphs in Figure~\ref{fig:framework}(c)  and similarly adds time embedding as input $\mathbf{e}_m$, which is devised to validate whether our period-varying modeling method is better than the method of simply adding period information.
\end{itemize}

To evaluate the improvement of our model in different periods, Figure ~\ref{fig:small_time} and Figure ~\ref{fig:big_time} present the performance of the competitive baselines SVD-GCN, DPVP as well as the above-mentioned three ablation baselines on the two datasets. Details of performance on AUC metric for the two datasets can be found in Appendix~\ref{appendix:impact_supplement}.
Please note that the \textit{Full} in figures represents the performance of models for full-period test data without distinguishing the time periods. According to the results, we can observe that:
\begin{itemize}
\item Encouragingly, DPVP achieves the best results under various metrics over different time periods.
\item With time embedding as input, the performance of SVD-GCN(time emb) and DPVP(time emb) is improved compared to the corresponding baselines (i.e., SVD-GCN and DPVP(w/o time)) ignoring the time information, which verifies that time information is beneficial to the modeling of takeaway recommendation indeed.
\item Compared to DPVP(w/o time) without period-varying modeling component and DPVP(time emb)  simply taking time embedding as input, DPVP performs better through modeling subgraphs of different periods constructed based on the periodic patterns. 
Specifically, compared with DPVP(w/o time), for MT-large dataset,  the minimum improvement of the three metrics  over different periods is 1.17\% (i.e., MRR in Noon period) and for MT-small dataset, the smallest improvement of these metrics is 3.91\% (i.e., Hit@10 for Late Night period).
\end{itemize}

In particular, to prove our model can effectively capture the periodic period-varying patterns shown in Figure~\ref{fig:distribution_time_interation}, we provide a visualization of the average predicted scores (i.e., $\hat{y}_{u,s,{O},m} $ in Eq.~\ref{eq:predict}) of five major food categories o (o in ${O}$) over different time periods in MT-small test data.
As shown in Figure ~\ref{fig:heat}, we can clearly observe that:   the predicted score of Baozi/Porridge commonly served as a breakfast food in the Morning period is significantly higher than that in other periods; Likewise,  the barbecue, which is usually chosen as a midnight snack, also scores higher in the Late Night period; Besides, encouragingly, the predicted scores of Dish are higher in Noon and Night periods than those in Morning and Late Night periods as expected; Moreover, the interaction characteristics between users and noodle category do not change significantly over time, as shown in Figure ~\ref{fig:distribution_time_interation}. Within the expectation, there is no apparent difference in the predicted scores of noodle over different time periods.



\subsection{Online A/B test(RQ3)}
We deploy our method to the Meituan Takeaway recommender system and conduct a two-week online A/B test. 
We leverage the learned graph embeddings from our DPVP model as additional features in the downstream recommendation model, and increase the GMV(Gross Merchandise Volume) by 0.70\%, the CTCVR(Click Through \& Conversion Rate)~\cite{ma2018entire} by 0.68\% in one of the main recommendation scenarios, which demonstrates the effectiveness of our method.  

\subsection{Hyper-Parameter Studies (RQ4)}\label{sec:hyper_parameter}
\textbf{Impact of Model Depth}.
To study the impact of the model depth of the graphs, we vary $L$ from 1 to 4. As shown in Figure ~\ref{fig:layer_ab} shown, we find that DPVP basically improves first and then drops, which is a trade-off of performance gain from high-order information (by stacking $L$ layers to reach $L$-hop neighbors) and performance degradation caused by over-smoothing. Since DPVP achieves the best performance with two layers, we set $L=2$ for both datasets.

\noindent\textbf{Impact of Embedding Dimension}.
To investigate the performance of DPVP on the two datasets with respect to the dimension of embedding vectors (i.e., $d$), we vary $d$ in $\{20,50,100,200\}$. As shown in Figure ~\ref{fig:dim_ab}, with the increase of embedding dimension, the performance of DPVP shows a trend of first increasing and then decreasing and DPVP performs best with the embedding dimension equaling $100$. Thus, we set $d=100$ for the two datasets.

\section{Conclusion}
In this paper, in order to model users' dual period-varying preferences in takeaway recommendation, we propose a novel graph-based model, named DPVP. Specifically, we design a dual interaction-aware module for modeling the dual preferences. Furthermore, we propose a time-based decomposition module as well as a time-aware gating network for capturing time-varying preferences. Extensive experiments on two real-world datasets from Meituan platform and online A/B tests empirically demonstrate the superiority of our model. Additionally, studies prove that our model can improve the recommendation performance for all time periods and is able to capture users' varying preferences over different periods. As further work, we plan to explore incorporating additional time factors, such as seasons and holidays, to more accurately model users' varying preferences for takeaway recommendation.

\begin{acks}
The research work is supported by the National Key Research and Development Program of China under Grant No. 2021ZD0113602, the National Natural Science Foundation of China under Grant No. 62176014, the Fundamental Research Funds for the Central Universities.
\end{acks}
\clearpage
\bibliographystyle{ACM-Reference-Format}
\balance
\bibliography{sample-sigconf}


\begin{thebibliography}{45}


\ifx \showCODEN    \undefined \def \showCODEN     #1{\unskip}     \fi
\ifx \showDOI      \undefined \def \showDOI       #1{#1}\fi
\ifx \showISBNx    \undefined \def \showISBNx     #1{\unskip}     \fi
\ifx \showISBNxiii \undefined \def \showISBNxiii  #1{\unskip}     \fi
\ifx \showISSN     \undefined \def \showISSN      #1{\unskip}     \fi
\ifx \showLCCN     \undefined \def \showLCCN      #1{\unskip}     \fi
\ifx \shownote     \undefined \def \shownote      #1{#1}          \fi
\ifx \showarticletitle \undefined \def \showarticletitle #1{#1}   \fi
\ifx \showURL      \undefined \def \showURL       {\relax}        \fi
\providecommand\bibfield[2]{#2}
\providecommand\bibinfo[2]{#2}
\providecommand\natexlab[1]{#1}
\providecommand\showeprint[2][]{arXiv:#2}

\bibitem[now(2023)]%
        {now_statistic}
 \bibinfo{year}{2023}\natexlab{}.
\newblock \bibinfo{title}{Number of users of the online food delivery market
  worldwide in 2022 and 2027, by region}.
\newblock
  \bibinfo{howpublished}{\url{https://www.statista.com/forecasts/1358171/online-food-delivery-users-by-region-worldwide}}.
\newblock


\bibitem[Berg et~al\mbox{.}(2017)]%
        {berg2017graph}
\bibfield{author}{\bibinfo{person}{Rianne van~den Berg},
  \bibinfo{person}{Thomas~N Kipf}, {and} \bibinfo{person}{Max Welling}.}
  \bibinfo{year}{2017}\natexlab{}.
\newblock \showarticletitle{Graph convolutional matrix completion}.
\newblock \bibinfo{journal}{\emph{arXiv preprint arXiv:1706.02263}}
  (\bibinfo{year}{2017}).
\newblock


\bibitem[Chen et~al\mbox{.}(2018)]%
        {chen2018neural}
\bibfield{author}{\bibinfo{person}{Chong Chen}, \bibinfo{person}{Min Zhang},
  \bibinfo{person}{Yiqun Liu}, {and} \bibinfo{person}{Shaoping Ma}.}
  \bibinfo{year}{2018}\natexlab{}.
\newblock \showarticletitle{Neural attentional rating regression with
  review-level explanations}. In \bibinfo{booktitle}{\emph{Proceedings of the
  2018 World Wide Web Conference}}. \bibinfo{pages}{1583--1592}.
\newblock


\bibitem[Chen et~al\mbox{.}(2020)]%
        {chen2020efficient}
\bibfield{author}{\bibinfo{person}{Chong Chen}, \bibinfo{person}{Min Zhang},
  \bibinfo{person}{Yongfeng Zhang}, \bibinfo{person}{Yiqun Liu}, {and}
  \bibinfo{person}{Shaoping Ma}.} \bibinfo{year}{2020}\natexlab{}.
\newblock \showarticletitle{Efficient neural matrix factorization without
  sampling for recommendation}.
\newblock \bibinfo{journal}{\emph{ACM Transactions on Information Systems
  (TOIS)}} \bibinfo{volume}{38}, \bibinfo{number}{2} (\bibinfo{year}{2020}),
  \bibinfo{pages}{1--28}.
\newblock


\bibitem[Eksombatchai et~al\mbox{.}(2018)]%
        {eksombatchai2018pixie}
\bibfield{author}{\bibinfo{person}{Chantat Eksombatchai},
  \bibinfo{person}{Pranav Jindal}, \bibinfo{person}{Jerry~Zitao Liu},
  \bibinfo{person}{Yuchen Liu}, \bibinfo{person}{Rahul Sharma},
  \bibinfo{person}{Charles Sugnet}, \bibinfo{person}{Mark Ulrich}, {and}
  \bibinfo{person}{Jure Leskovec}.} \bibinfo{year}{2018}\natexlab{}.
\newblock \showarticletitle{Pixie: A system for recommending 3+ billion items
  to 200+ million users in real-time}. In \bibinfo{booktitle}{\emph{Proceedings
  of the 2018 world wide web conference}}. \bibinfo{pages}{1775--1784}.
\newblock


\bibitem[Ekstrand et~al\mbox{.}(2011)]%
        {ekstrand2011collaborative}
\bibfield{author}{\bibinfo{person}{Michael~D Ekstrand}, \bibinfo{person}{John~T
  Riedl}, \bibinfo{person}{Joseph~A Konstan}, {et~al\mbox{.}}}
  \bibinfo{year}{2011}\natexlab{}.
\newblock \showarticletitle{Collaborative filtering recommender systems}.
\newblock \bibinfo{journal}{\emph{Foundations and Trends{\textregistered} in
  Human--Computer Interaction}} \bibinfo{volume}{4}, \bibinfo{number}{2}
  (\bibinfo{year}{2011}), \bibinfo{pages}{81--173}.
\newblock


\bibitem[Ferri et~al\mbox{.}(2011)]%
        {ferri2011coherent}
\bibfield{author}{\bibinfo{person}{Cesar Ferri}, \bibinfo{person}{Jos{\'e}
  Hern{\'a}ndez-Orallo}, {and} \bibinfo{person}{Peter~A Flach}.}
  \bibinfo{year}{2011}\natexlab{}.
\newblock \showarticletitle{A coherent interpretation of AUC as a measure of
  aggregated classification performance}. In
  \bibinfo{booktitle}{\emph{Proceedings of the 28th International Conference on
  Machine Learning (ICML-11)}}. \bibinfo{pages}{657--664}.
\newblock


\bibitem[Hamilton et~al\mbox{.}(2017)]%
        {hamilton2017inductive}
\bibfield{author}{\bibinfo{person}{Will Hamilton}, \bibinfo{person}{Zhitao
  Ying}, {and} \bibinfo{person}{Jure Leskovec}.}
  \bibinfo{year}{2017}\natexlab{}.
\newblock \showarticletitle{Inductive representation learning on large graphs}.
\newblock \bibinfo{journal}{\emph{Advances in neural information processing
  systems}}  \bibinfo{volume}{30} (\bibinfo{year}{2017}).
\newblock


\bibitem[He et~al\mbox{.}(2020)]%
        {he2020lightgcn}
\bibfield{author}{\bibinfo{person}{Xiangnan He}, \bibinfo{person}{Kuan Deng},
  \bibinfo{person}{Xiang Wang}, \bibinfo{person}{Yan Li},
  \bibinfo{person}{Yongdong Zhang}, {and} \bibinfo{person}{Meng Wang}.}
  \bibinfo{year}{2020}\natexlab{}.
\newblock \showarticletitle{Lightgcn: Simplifying and powering graph
  convolution network for recommendation}. In
  \bibinfo{booktitle}{\emph{Proceedings of the 43rd International ACM SIGIR
  conference on research and development in Information Retrieval}}.
  \bibinfo{pages}{639--648}.
\newblock


\bibitem[He et~al\mbox{.}(2017)]%
        {he2017neural}
\bibfield{author}{\bibinfo{person}{Xiangnan He}, \bibinfo{person}{Lizi Liao},
  \bibinfo{person}{Hanwang Zhang}, \bibinfo{person}{Liqiang Nie},
  \bibinfo{person}{Xia Hu}, {and} \bibinfo{person}{Tat-Seng Chua}.}
  \bibinfo{year}{2017}\natexlab{}.
\newblock \showarticletitle{Neural collaborative filtering}. In
  \bibinfo{booktitle}{\emph{Proceedings of the 26th international conference on
  world wide web}}. \bibinfo{pages}{173--182}.
\newblock


\bibitem[Hu et~al\mbox{.}(2020)]%
        {hu2020heterogeneous}
\bibfield{author}{\bibinfo{person}{Ziniu Hu}, \bibinfo{person}{Yuxiao Dong},
  \bibinfo{person}{Kuansan Wang}, {and} \bibinfo{person}{Yizhou Sun}.}
  \bibinfo{year}{2020}\natexlab{}.
\newblock \showarticletitle{Heterogeneous graph transformer}. In
  \bibinfo{booktitle}{\emph{Proceedings of The Web Conference 2020}}.
  \bibinfo{pages}{2704--2710}.
\newblock


\bibitem[J{\"a}rvelin and Kek{\"a}l{\"a}inen(2002)]%
        {jarvelin2002cumulated}
\bibfield{author}{\bibinfo{person}{Kalervo J{\"a}rvelin} {and}
  \bibinfo{person}{Jaana Kek{\"a}l{\"a}inen}.} \bibinfo{year}{2002}\natexlab{}.
\newblock \showarticletitle{Cumulated gain-based evaluation of IR techniques}.
\newblock \bibinfo{journal}{\emph{ACM Transactions on Information Systems
  (TOIS)}} \bibinfo{volume}{20}, \bibinfo{number}{4} (\bibinfo{year}{2002}),
  \bibinfo{pages}{422--446}.
\newblock


\bibitem[Kipf and Welling(2016)]%
        {kipf2016semi}
\bibfield{author}{\bibinfo{person}{Thomas~N Kipf} {and} \bibinfo{person}{Max
  Welling}.} \bibinfo{year}{2016}\natexlab{}.
\newblock \showarticletitle{Semi-supervised classification with graph
  convolutional networks}.
\newblock \bibinfo{journal}{\emph{arXiv preprint arXiv:1609.02907}}
  (\bibinfo{year}{2016}).
\newblock


\bibitem[Koren et~al\mbox{.}(2009)]%
        {koren2009matrix}
\bibfield{author}{\bibinfo{person}{Yehuda Koren}, \bibinfo{person}{Robert
  Bell}, {and} \bibinfo{person}{Chris Volinsky}.}
  \bibinfo{year}{2009}\natexlab{}.
\newblock \showarticletitle{Matrix factorization techniques for recommender
  systems}.
\newblock \bibinfo{journal}{\emph{Computer}} \bibinfo{volume}{42},
  \bibinfo{number}{8} (\bibinfo{year}{2009}), \bibinfo{pages}{30--37}.
\newblock


\bibitem[Li et~al\mbox{.}(2022)]%
        {li2022automatically}
\bibfield{author}{\bibinfo{person}{Yinfeng Li}, \bibinfo{person}{Chen Gao},
  \bibinfo{person}{Xiaoyi Du}, \bibinfo{person}{Huazhou Wei},
  \bibinfo{person}{Hengliang Luo}, \bibinfo{person}{Depeng Jin}, {and}
  \bibinfo{person}{Yong Li}.} \bibinfo{year}{2022}\natexlab{}.
\newblock \showarticletitle{Automatically Discovering User Consumption Intents
  in Meituan}. In \bibinfo{booktitle}{\emph{Proceedings of the 28th ACM SIGKDD
  Conference on Knowledge Discovery and Data Mining}}.
  \bibinfo{pages}{3259--3269}.
\newblock


\bibitem[Lin et~al\mbox{.}(2022)]%
        {lin2022spatiotemporal}
\bibfield{author}{\bibinfo{person}{Shaochuan Lin}, \bibinfo{person}{Yicong Yu},
  \bibinfo{person}{Xiyu Ji}, \bibinfo{person}{Taotao Zhou},
  \bibinfo{person}{Hengxu He}, \bibinfo{person}{Zisen Sang},
  \bibinfo{person}{Jia Jia}, \bibinfo{person}{Guodong Cao}, {and}
  \bibinfo{person}{Ning Hu}.} \bibinfo{year}{2022}\natexlab{}.
\newblock \showarticletitle{Spatiotemporal-Enhanced Network for Click-Through
  Rate Prediction in Location-based Services}.
\newblock \bibinfo{journal}{\emph{arXiv preprint arXiv:2209.09427}}
  (\bibinfo{year}{2022}).
\newblock


\bibitem[Loshchilov and Hutter(2018)]%
        {loshchilov2018decoupled}
\bibfield{author}{\bibinfo{person}{Ilya Loshchilov} {and}
  \bibinfo{person}{Frank Hutter}.} \bibinfo{year}{2018}\natexlab{}.
\newblock \showarticletitle{Decoupled Weight Decay Regularization}. In
  \bibinfo{booktitle}{\emph{International Conference on Learning
  Representations}}.
\newblock


\bibitem[Ma et~al\mbox{.}(2018)]%
        {ma2018entire}
\bibfield{author}{\bibinfo{person}{Xiao Ma}, \bibinfo{person}{Liqin Zhao},
  \bibinfo{person}{Guan Huang}, \bibinfo{person}{Zhi Wang},
  \bibinfo{person}{Zelin Hu}, \bibinfo{person}{Xiaoqiang Zhu}, {and}
  \bibinfo{person}{Kun Gai}.} \bibinfo{year}{2018}\natexlab{}.
\newblock \showarticletitle{Entire space multi-task model: An effective
  approach for estimating post-click conversion rate}. In
  \bibinfo{booktitle}{\emph{The 41st International ACM SIGIR Conference on
  Research \& Development in Information Retrieval}}.
  \bibinfo{pages}{1137--1140}.
\newblock


\bibitem[Mao et~al\mbox{.}(2021a)]%
        {mao2021simplex}
\bibfield{author}{\bibinfo{person}{Kelong Mao}, \bibinfo{person}{Jieming Zhu},
  \bibinfo{person}{Jinpeng Wang}, \bibinfo{person}{Quanyu Dai},
  \bibinfo{person}{Zhenhua Dong}, \bibinfo{person}{Xi Xiao}, {and}
  \bibinfo{person}{Xiuqiang He}.} \bibinfo{year}{2021}\natexlab{a}.
\newblock \showarticletitle{SimpleX: A simple and strong baseline for
  collaborative filtering}. In \bibinfo{booktitle}{\emph{Proceedings of the
  30th ACM International Conference on Information \& Knowledge Management}}.
  \bibinfo{pages}{1243--1252}.
\newblock


\bibitem[Mao et~al\mbox{.}(2021b)]%
        {mao2021ultragcn}
\bibfield{author}{\bibinfo{person}{Kelong Mao}, \bibinfo{person}{Jieming Zhu},
  \bibinfo{person}{Xi Xiao}, \bibinfo{person}{Biao Lu},
  \bibinfo{person}{Zhaowei Wang}, {and} \bibinfo{person}{Xiuqiang He}.}
  \bibinfo{year}{2021}\natexlab{b}.
\newblock \showarticletitle{UltraGCN: ultra simplification of graph
  convolutional networks for recommendation}. In
  \bibinfo{booktitle}{\emph{Proceedings of the 30th ACM International
  Conference on Information \& Knowledge Management}}.
  \bibinfo{pages}{1253--1262}.
\newblock


\bibitem[Peng et~al\mbox{.}(2022)]%
        {peng2022svd}
\bibfield{author}{\bibinfo{person}{Shaowen Peng}, \bibinfo{person}{Kazunari
  Sugiyama}, {and} \bibinfo{person}{Tsunenori Mine}.}
  \bibinfo{year}{2022}\natexlab{}.
\newblock \showarticletitle{SVD-GCN: A Simplified Graph Convolution Paradigm
  for Recommendation}. In \bibinfo{booktitle}{\emph{Proceedings of the 31st ACM
  International Conference on Information \& Knowledge Management}}.
  \bibinfo{pages}{1625--1634}.
\newblock


\bibitem[Pfadler et~al\mbox{.}(2020)]%
        {pfadler2020billion}
\bibfield{author}{\bibinfo{person}{Andreas Pfadler}, \bibinfo{person}{Huan
  Zhao}, \bibinfo{person}{Jizhe Wang}, \bibinfo{person}{Lifeng Wang},
  \bibinfo{person}{Pipei Huang}, {and} \bibinfo{person}{Dik~Lun Lee}.}
  \bibinfo{year}{2020}\natexlab{}.
\newblock \showarticletitle{Billion-scale recommendation with heterogeneous
  side information at taobao}. In \bibinfo{booktitle}{\emph{2020 IEEE 36th
  International Conference on Data Engineering (ICDE)}}. IEEE,
  \bibinfo{pages}{1667--1676}.
\newblock


\bibitem[Qi et~al\mbox{.}(2021)]%
        {qi2021trilateral}
\bibfield{author}{\bibinfo{person}{Yi Qi}, \bibinfo{person}{Ke Hu},
  \bibinfo{person}{Bo Zhang}, \bibinfo{person}{Jia Cheng}, {and}
  \bibinfo{person}{Jun Lei}.} \bibinfo{year}{2021}\natexlab{}.
\newblock \showarticletitle{Trilateral Spatiotemporal Attention Network for
  User Behavior Modeling in Location-based Search}. In
  \bibinfo{booktitle}{\emph{Proceedings of the 30th ACM International
  Conference on Information \& Knowledge Management}}.
  \bibinfo{pages}{3373--3377}.
\newblock


\bibitem[Radev et~al\mbox{.}(2002)]%
        {radev2002evaluating}
\bibfield{author}{\bibinfo{person}{Dragomir~R Radev}, \bibinfo{person}{Hong
  Qi}, \bibinfo{person}{Harris Wu}, {and} \bibinfo{person}{Weiguo Fan}.}
  \bibinfo{year}{2002}\natexlab{}.
\newblock \showarticletitle{Evaluating Web-based Question Answering Systems.}.
  In \bibinfo{booktitle}{\emph{LREC}}. Citeseer.
\newblock


\bibitem[Rendle et~al\mbox{.}(2009)]%
        {rendle2009bpr}
\bibfield{author}{\bibinfo{person}{Steffen Rendle}, \bibinfo{person}{Christoph
  Freudenthaler}, \bibinfo{person}{Zeno Gantner}, {and} \bibinfo{person}{Lars
  Schmidt-Thieme}.} \bibinfo{year}{2009}\natexlab{}.
\newblock \showarticletitle{BPR: Bayesian personalized ranking from implicit
  feedback}. In \bibinfo{booktitle}{\emph{Proceedings of the Twenty-Fifth
  Conference on Uncertainty in Artificial Intelligence}}.
  \bibinfo{pages}{452--461}.
\newblock


\bibitem[Scarselli et~al\mbox{.}(2008)]%
        {scarselli2008graph}
\bibfield{author}{\bibinfo{person}{Franco Scarselli}, \bibinfo{person}{Marco
  Gori}, \bibinfo{person}{Ah~Chung Tsoi}, \bibinfo{person}{Markus
  Hagenbuchner}, {and} \bibinfo{person}{Gabriele Monfardini}.}
  \bibinfo{year}{2008}\natexlab{}.
\newblock \showarticletitle{The graph neural network model}.
\newblock \bibinfo{journal}{\emph{IEEE transactions on neural networks}}
  \bibinfo{volume}{20}, \bibinfo{number}{1} (\bibinfo{year}{2008}),
  \bibinfo{pages}{61--80}.
\newblock


\bibitem[Seo et~al\mbox{.}(2017)]%
        {seo2017representation}
\bibfield{author}{\bibinfo{person}{Sungyong Seo}, \bibinfo{person}{Jing Huang},
  \bibinfo{person}{Hao Yang}, {and} \bibinfo{person}{Yan Liu}.}
  \bibinfo{year}{2017}\natexlab{}.
\newblock \showarticletitle{Representation learning of users and items for
  review rating prediction using attention-based convolutional neural network}.
  In \bibinfo{booktitle}{\emph{International Workshop on Machine Learning
  Methods for Recommender Systems}}.
\newblock


\bibitem[Sun et~al\mbox{.}(2019)]%
        {sun2019multi}
\bibfield{author}{\bibinfo{person}{Jianing Sun}, \bibinfo{person}{Yingxue
  Zhang}, \bibinfo{person}{Chen Ma}, \bibinfo{person}{Mark Coates},
  \bibinfo{person}{Huifeng Guo}, \bibinfo{person}{Ruiming Tang}, {and}
  \bibinfo{person}{Xiuqiang He}.} \bibinfo{year}{2019}\natexlab{}.
\newblock \showarticletitle{Multi-graph convolution collaborative filtering}.
  In \bibinfo{booktitle}{\emph{2019 IEEE international conference on data
  mining (ICDM)}}. IEEE, \bibinfo{pages}{1306--1311}.
\newblock


\bibitem[Vaswani et~al\mbox{.}(2017)]%
        {vaswani2017attention}
\bibfield{author}{\bibinfo{person}{Ashish Vaswani}, \bibinfo{person}{Noam
  Shazeer}, \bibinfo{person}{Niki Parmar}, \bibinfo{person}{Jakob Uszkoreit},
  \bibinfo{person}{Llion Jones}, \bibinfo{person}{Aidan~N Gomez},
  \bibinfo{person}{{\L}ukasz Kaiser}, {and} \bibinfo{person}{Illia
  Polosukhin}.} \bibinfo{year}{2017}\natexlab{}.
\newblock \showarticletitle{Attention is all you need}.
\newblock \bibinfo{journal}{\emph{Advances in neural information processing
  systems}}  \bibinfo{volume}{30} (\bibinfo{year}{2017}).
\newblock


\bibitem[Veli{\v{c}}kovi{\'c} et~al\mbox{.}(2017)]%
        {velivckovic2017graph}
\bibfield{author}{\bibinfo{person}{Petar Veli{\v{c}}kovi{\'c}},
  \bibinfo{person}{Guillem Cucurull}, \bibinfo{person}{Arantxa Casanova},
  \bibinfo{person}{Adriana Romero}, \bibinfo{person}{Pietro Lio}, {and}
  \bibinfo{person}{Yoshua Bengio}.} \bibinfo{year}{2017}\natexlab{}.
\newblock \showarticletitle{Graph attention networks}.
\newblock \bibinfo{journal}{\emph{arXiv preprint arXiv:1710.10903}}
  (\bibinfo{year}{2017}).
\newblock


\bibitem[Wang et~al\mbox{.}(2020)]%
        {wang2020calendar}
\bibfield{author}{\bibinfo{person}{Daheng Wang}, \bibinfo{person}{Meng Jiang},
  \bibinfo{person}{Munira Syed}, \bibinfo{person}{Oliver Conway},
  \bibinfo{person}{Vishal Juneja}, \bibinfo{person}{Sriram Subramanian}, {and}
  \bibinfo{person}{Nitesh~V Chawla}.} \bibinfo{year}{2020}\natexlab{}.
\newblock \showarticletitle{Calendar graph neural networks for modeling time
  structures in spatiotemporal user behaviors}. In
  \bibinfo{booktitle}{\emph{Proceedings of the 26th ACM SIGKDD international
  conference on knowledge discovery \& data mining}}.
  \bibinfo{pages}{2581--2589}.
\newblock


\bibitem[Wang et~al\mbox{.}(2019)]%
        {wang2019neural}
\bibfield{author}{\bibinfo{person}{Xiang Wang}, \bibinfo{person}{Xiangnan He},
  \bibinfo{person}{Meng Wang}, \bibinfo{person}{Fuli Feng}, {and}
  \bibinfo{person}{Tat-Seng Chua}.} \bibinfo{year}{2019}\natexlab{}.
\newblock \showarticletitle{Neural graph collaborative filtering}. In
  \bibinfo{booktitle}{\emph{Proceedings of the 42nd international ACM SIGIR
  conference on Research and development in Information Retrieval}}.
  \bibinfo{pages}{165--174}.
\newblock


\bibitem[Wu et~al\mbox{.}(2019a)]%
        {wu2019simplifying}
\bibfield{author}{\bibinfo{person}{Felix Wu}, \bibinfo{person}{Amauri Souza},
  \bibinfo{person}{Tianyi Zhang}, \bibinfo{person}{Christopher Fifty},
  \bibinfo{person}{Tao Yu}, {and} \bibinfo{person}{Kilian Weinberger}.}
  \bibinfo{year}{2019}\natexlab{a}.
\newblock \showarticletitle{Simplifying graph convolutional networks}. In
  \bibinfo{booktitle}{\emph{International conference on machine learning}}.
  PMLR, \bibinfo{pages}{6861--6871}.
\newblock


\bibitem[Wu et~al\mbox{.}(2019b)]%
        {wu2019dual}
\bibfield{author}{\bibinfo{person}{Qitian Wu}, \bibinfo{person}{Hengrui Zhang},
  \bibinfo{person}{Xiaofeng Gao}, \bibinfo{person}{Peng He},
  \bibinfo{person}{Paul Weng}, \bibinfo{person}{Han Gao}, {and}
  \bibinfo{person}{Guihai Chen}.} \bibinfo{year}{2019}\natexlab{b}.
\newblock \showarticletitle{Dual graph attention networks for deep latent
  representation of multifaceted social effects in recommender systems}. In
  \bibinfo{booktitle}{\emph{The world wide web conference}}.
  \bibinfo{pages}{2091--2102}.
\newblock


\bibitem[Wu et~al\mbox{.}(2022)]%
        {wu2022graph}
\bibfield{author}{\bibinfo{person}{Shiwen Wu}, \bibinfo{person}{Fei Sun},
  \bibinfo{person}{Wentao Zhang}, \bibinfo{person}{Xu Xie}, {and}
  \bibinfo{person}{Bin Cui}.} \bibinfo{year}{2022}\natexlab{}.
\newblock \showarticletitle{Graph neural networks in recommender systems: a
  survey}.
\newblock \bibinfo{journal}{\emph{Comput. Surveys}} \bibinfo{volume}{55},
  \bibinfo{number}{5} (\bibinfo{year}{2022}), \bibinfo{pages}{1--37}.
\newblock


\bibitem[Wu et~al\mbox{.}(2020)]%
        {wu2020comprehensive}
\bibfield{author}{\bibinfo{person}{Zonghan Wu}, \bibinfo{person}{Shirui Pan},
  \bibinfo{person}{Fengwen Chen}, \bibinfo{person}{Guodong Long},
  \bibinfo{person}{Chengqi Zhang}, {and} \bibinfo{person}{S~Yu Philip}.}
  \bibinfo{year}{2020}\natexlab{}.
\newblock \showarticletitle{A comprehensive survey on graph neural networks}.
\newblock \bibinfo{journal}{\emph{IEEE transactions on neural networks and
  learning systems}} \bibinfo{volume}{32}, \bibinfo{number}{1}
  (\bibinfo{year}{2020}), \bibinfo{pages}{4--24}.
\newblock


\bibitem[Xu et~al\mbox{.}(2020)]%
        {xu2020graphsail}
\bibfield{author}{\bibinfo{person}{Yishi Xu}, \bibinfo{person}{Yingxue Zhang},
  \bibinfo{person}{Wei Guo}, \bibinfo{person}{Huifeng Guo},
  \bibinfo{person}{Ruiming Tang}, {and} \bibinfo{person}{Mark Coates}.}
  \bibinfo{year}{2020}\natexlab{}.
\newblock \showarticletitle{Graphsail: Graph structure aware incremental
  learning for recommender systems}. In \bibinfo{booktitle}{\emph{Proceedings
  of the 29th ACM International Conference on Information \& Knowledge
  Management}}. \bibinfo{pages}{2861--2868}.
\newblock


\bibitem[Yang et~al\mbox{.}(2022)]%
        {yang2022gated}
\bibfield{author}{\bibinfo{person}{Tianchi Yang}, \bibinfo{person}{Luhao
  Zhang}, \bibinfo{person}{Chuan Shi}, \bibinfo{person}{Cheng Yang},
  \bibinfo{person}{Siyong Xu}, \bibinfo{person}{Ruiyu Fang},
  \bibinfo{person}{Maodi Hu}, \bibinfo{person}{Huaijun Liu},
  \bibinfo{person}{Tao Li}, {and} \bibinfo{person}{Dong Wang}.}
  \bibinfo{year}{2022}\natexlab{}.
\newblock \showarticletitle{Gated Hypergraph Neural Network for Scene-Aware
  Recommendation}. In \bibinfo{booktitle}{\emph{International Conference on
  Database Systems for Advanced Applications}}. Springer,
  \bibinfo{pages}{199--215}.
\newblock


\bibitem[Ying et~al\mbox{.}(2018)]%
        {ying2018graph}
\bibfield{author}{\bibinfo{person}{Rex Ying}, \bibinfo{person}{Ruining He},
  \bibinfo{person}{Kaifeng Chen}, \bibinfo{person}{Pong Eksombatchai},
  \bibinfo{person}{William~L Hamilton}, {and} \bibinfo{person}{Jure Leskovec}.}
  \bibinfo{year}{2018}\natexlab{}.
\newblock \showarticletitle{Graph convolutional neural networks for web-scale
  recommender systems}. In \bibinfo{booktitle}{\emph{Proceedings of the 24th
  ACM SIGKDD international conference on knowledge discovery \& data mining}}.
  \bibinfo{pages}{974--983}.
\newblock


\bibitem[Yu et~al\mbox{.}(2021)]%
        {yu2021dual}
\bibfield{author}{\bibinfo{person}{Yantao Yu}, \bibinfo{person}{Weipeng Wang},
  \bibinfo{person}{Zhoutian Feng}, {and} \bibinfo{person}{Daiyue Xue}.}
  \bibinfo{year}{2021}\natexlab{}.
\newblock \showarticletitle{A dual augmented two-tower model for online
  large-scale recommendation}.
\newblock \bibinfo{journal}{\emph{DLP-KDD}} (\bibinfo{year}{2021}).
\newblock


\bibitem[Zhang et~al\mbox{.}(2019)]%
        {zhang2019deep}
\bibfield{author}{\bibinfo{person}{Shuai Zhang}, \bibinfo{person}{Lina Yao},
  \bibinfo{person}{Aixin Sun}, {and} \bibinfo{person}{Yi Tay}.}
  \bibinfo{year}{2019}\natexlab{}.
\newblock \showarticletitle{Deep learning based recommender system: A survey
  and new perspectives}.
\newblock \bibinfo{journal}{\emph{ACM Computing Surveys (CSUR)}}
  \bibinfo{volume}{52}, \bibinfo{number}{1} (\bibinfo{year}{2019}),
  \bibinfo{pages}{1--38}.
\newblock


\bibitem[Zhang et~al\mbox{.}(2020)]%
        {zhang2020deep}
\bibfield{author}{\bibinfo{person}{Ziwei Zhang}, \bibinfo{person}{Peng Cui},
  {and} \bibinfo{person}{Wenwu Zhu}.} \bibinfo{year}{2020}\natexlab{}.
\newblock \showarticletitle{Deep learning on graphs: A survey}.
\newblock \bibinfo{journal}{\emph{IEEE Transactions on Knowledge and Data
  Engineering}} (\bibinfo{year}{2020}).
\newblock


\bibitem[Zhao et~al\mbox{.}(2019)]%
        {zhao2019intentgc}
\bibfield{author}{\bibinfo{person}{Jun Zhao}, \bibinfo{person}{Zhou Zhou},
  \bibinfo{person}{Ziyu Guan}, \bibinfo{person}{Wei Zhao}, \bibinfo{person}{Wei
  Ning}, \bibinfo{person}{Guang Qiu}, {and} \bibinfo{person}{Xiaofei He}.}
  \bibinfo{year}{2019}\natexlab{}.
\newblock \showarticletitle{Intentgc: a scalable graph convolution framework
  fusing heterogeneous information for recommendation}. In
  \bibinfo{booktitle}{\emph{Proceedings of the 25th ACM SIGKDD International
  Conference on Knowledge Discovery \& Data Mining}}.
  \bibinfo{pages}{2347--2357}.
\newblock


\bibitem[Zheng et~al\mbox{.}(2017)]%
        {zheng2017joint}
\bibfield{author}{\bibinfo{person}{Lei Zheng}, \bibinfo{person}{Vahid Noroozi},
  {and} \bibinfo{person}{Philip~S Yu}.} \bibinfo{year}{2017}\natexlab{}.
\newblock \showarticletitle{Joint deep modeling of users and items using
  reviews for recommendation}. In \bibinfo{booktitle}{\emph{Proceedings of the
  tenth ACM international conference on web search and data mining}}.
  \bibinfo{pages}{425--434}.
\newblock


\bibitem[Zhou et~al\mbox{.}(2020)]%
        {zhou2020graph}
\bibfield{author}{\bibinfo{person}{Jie Zhou}, \bibinfo{person}{Ganqu Cui},
  \bibinfo{person}{Shengding Hu}, \bibinfo{person}{Zhengyan Zhang},
  \bibinfo{person}{Cheng Yang}, \bibinfo{person}{Zhiyuan Liu},
  \bibinfo{person}{Lifeng Wang}, \bibinfo{person}{Changcheng Li}, {and}
  \bibinfo{person}{Maosong Sun}.} \bibinfo{year}{2020}\natexlab{}.
\newblock \showarticletitle{Graph neural networks: A review of methods and
  applications}.
\newblock \bibinfo{journal}{\emph{AI Open}}  \bibinfo{volume}{1}
  (\bibinfo{year}{2020}), \bibinfo{pages}{57--81}.
\newblock


\end{thebibliography}
\clearpage
\appendix
\section{Experiment}
\subsection{Data desription}\label{appendix:data_description}
The details of the statistics of MT-small and MT-large datasets are summarized in Table \ref{tab:statistics}.  
\begin{table}[H]
\setlength{\abovecaptionskip}{0.1cm}
	\setlength{\belowcaptionskip}{-0.15cm}

  \caption{The statistics of the two datasets from Meituan.}
  \label{tab:statistics}
  \begin{tabular}{ccccc}
  \toprule
    Dataset & \#User & \#Store & \#Food &\#User-Store Interaction \\ 
    \midrule
    
    MT-small &56,887&4,059&5,952&180,283 \\
    MT-large&385,381&18,770&17,111&1,492,164\\
  \bottomrule
\end{tabular}
\end{table}
\subsection{Detailed Implementation}\label{appendix:implementation}
The hidden sizes of MLP described in Section~\ref{sec:prediction} are set as [400,200] by default.
For each store, we get its 10 most frequently clicked foods as its candidate food in the training data (i.e., $N'=10$ in Eq.~\ref{eq:food gate}).  Section~\ref{sec:hyper_parameter} reports the
impact of other essential hyper-parameters including the model depth $L$ and the dimension of latent vectors $d$ (i.e., the embeddings of users, stores, foods and time periods) in DPVP, and we utilize the best parameter settings. 
Under the above settings, all models are trained using AdamW optimizer \cite{loshchilov2018decoupled} with a learning rate of 1e-4, batch size of 1024 and weight decay of 1e-5.
\subsection{Baselines}\label{appendix:baselines}
We compare our proposed model against two groups of state-of-the-art recommendation methods as follows:
\\
(1) \textit{Graph-free} methods: 
\begin{itemize}
    \item NeuMF~\cite{he2017neural}. This method combines linear matrix factorization with the nonlinear MLP  to extract low-order and high-order features for prediction. 
    \item DNN. This is a vanilla deep-learning model for recommendation prediction.
    \item ENMF~\cite{chen2020efficient}. This model is a neural recommendation model that considers all samples in each parameter update without sampling.
    \item SimpleX~\cite{mao2021simplex}. This method integrates Matrix Factorization and user behavior modeling for prediction.
\end{itemize}
(2) \textit{Graph-based} methods: 
\begin{itemize}
    \item GCN~\cite{scarselli2008graph}. This method enrichs user and store representations through feature transformation, neighborhood aggregation, and nonlinear activation on graph. 
    
    \item GAT~\cite{velivckovic2017graph}. Considering the different importance of neighbor nodes to the central node, this model leverages the attention mechanism~\cite{vaswani2017attention} to weight the importance of neighbors in the aggregation process.
    \item NGCF~\cite{wang2019neural}. This model advances the vanilla GCN by additionally encoding the interactions via an element-wise multiplication. 
    \item HGT~\cite{hu2020heterogeneous}. This GNN model designs edge-type and node-type dependent parameters to deal with different interaction types and learn the node representations from the heterogeneous graph.
    \item LightGCN~\cite{he2020lightgcn}. This method simplifies GCN by removing unnecessary feature transformation and nonlinear activation parts. It merely adopts neighborhood aggregation to capture the collaborative filtering effect.

    \item UltraGCN~\cite{mao2021ultragcn}. This model further simplifies LightGCN by replacing neighborhood aggregation with weighted MF, leading to faster convergence and less complexity.
    \item SVD-GCN~\cite{peng2022svd}. This method simplifies GCN-based methods by solely exploiting K-largest singular vectors of a flexible truncated SVD  for the recommendation.  
\end{itemize}

All graph-based methods are aggregated on the \textit{full-period global graph} for the enriched representations. Since we focus on store recommendation in this work, we treat the store as the item. In all the  baselines except CF-based methods (i.e., NeuMF and NGCF) , we adopt MLP described in Section~\ref{sec:prediction} as the final prediction layer with user and store representations as input. To be fair, the parameter settings of MLP in baselines are the same as that of our proposed DPVP.

\subsection{Supplement for impact of period-varying modeling}\label{appendix:impact_supplement}
\begin{figure}[H]
    \centering
    \includegraphics[width=0.8\linewidth]{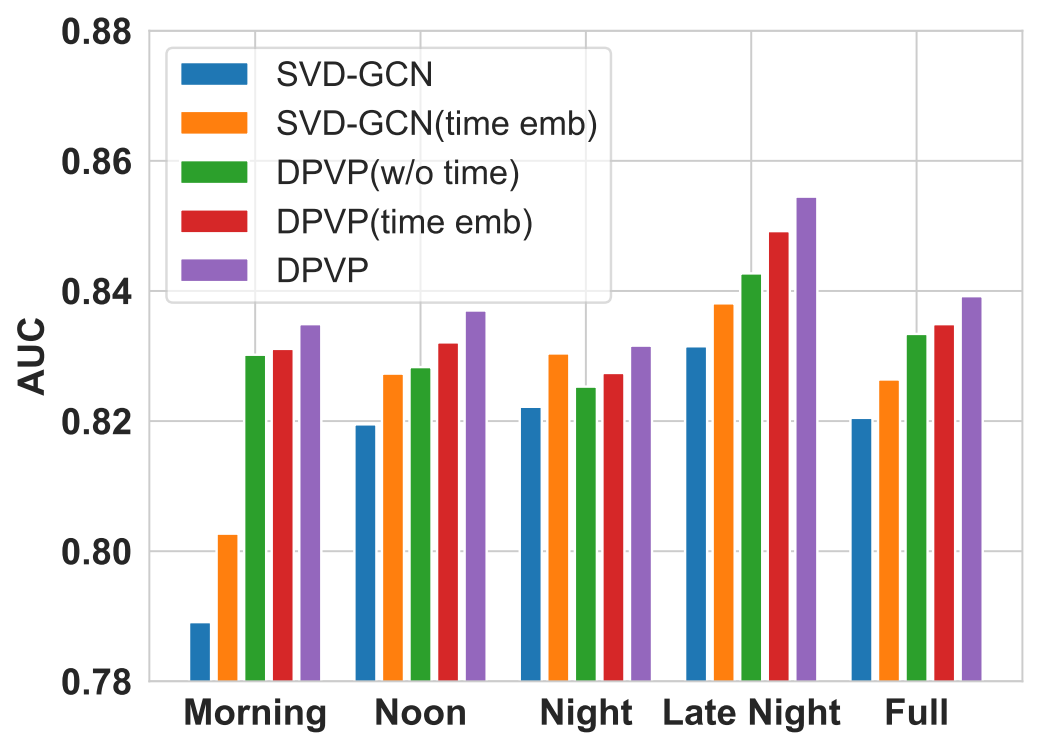}
    \caption{Impact of period-varying modeling on results for different time periods on AUC metric in MT-small dataset.}
    \label{fig:small_time_auc}
\end{figure}

\begin{figure}[H]
    \centering
    \includegraphics[width=0.8\linewidth]{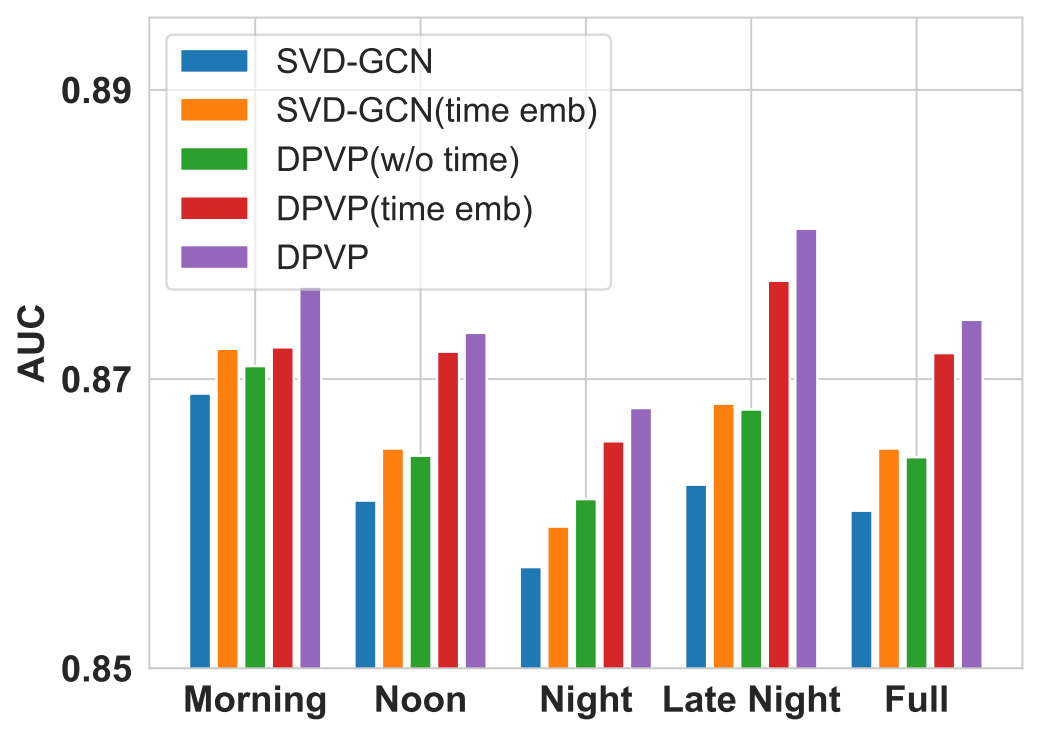}
    \caption{Impact of period-varying modeling on results for different time periods on AUC metric in MT-large dataset.}
    \label{fig:big_time_auc}
\end{figure}

In Figure ~\ref{fig:small_time_auc} and Figure ~\ref{fig:big_time_auc}, we present the impact of period-varying modeling on results for different time periods on AUC metric in the two datasets.


\end{document}